\documentclass[aps,showpacs,preprintnumbers,amsmath,amssymb,nofootinbib,eqsecnum, preprintnumbers]{revtex4}
\usepackage{graphicx}
\usepackage{epsf}
\usepackage{amsmath}
\usepackage{epstopdf}
\usepackage{bm}
\usepackage{color}
\usepackage{tabularx}
\usepackage{enumitem}
\usepackage{float}
\usepackage{array,booktabs}
\usepackage{footnote}
\usepackage{threeparttable}
\usepackage{graphicx}
\usepackage{hyperref}
\usepackage{amssymb,epsf}
\usepackage{latexsym}
\usepackage{epstopdf}
\usepackage{epsfig}
\usepackage{natbib}

\begin{document}

\title{Holographic Joule-Thomson Expansion in Lower Dimensions}
\author{Soodeh Zarepour$^{1}$\footnote{email address: szarepour@phys.usb.ac.ir}} \affiliation{$^1$
Department of Physics, University of Sistan and Baluchestan, Zahedan, Iran}

\begin{abstract}
The thermodynamical process of Joule-Thomson expansion is extended to the lower-dimensional spacetimes by considering rotating, charged and charged rotating BTZ black hole metrics. The Joule-Thomson coefficients, the inversion curves and the isenthalps are found for all three cases. From the behavior of inversion curves in the $T-P$ plane, it is observed that there exists only minimum inversion temperatures for BTZ black holes. The relevant calculations for rotating BTZ black holes are almost straightforward, while it is more challenging to address when a charge parameter is added. In the case of charged BTZ metrics, the black hole system can be considered in two possible thermodynamical schemes and it is explicitly proved that the Joule-Thomson expansion is present in a specific scheme (referred to as scheme II) in which thermodynamic instabilities do not arise. The existence of Joule-Thomson expansion in all three BTZ cases strongly indicates that these lower-dimensional black hole spacetimes can be considered as interacting statistical systems like their higher-dimensional cousins. 
\end{abstract}

\pacs{04.60.-M, 04.70.Dy} \maketitle

\section{Introduction}
Since the early days of black hole thermodynamics, when Beckenstein associated entropy with black holes \cite{beck1,beck2} and Hawking predicted the thermal emission of radiation for black holes \cite{hawk1,hawk2,hawk3}, the cosmological constant ($\Lambda$) has traditionally been served as a fixed parameter \cite{HawkingPage1983,Myers1988,Chamblin1,Chamblin2,Fernando}. Later on, it was suggested that the cosmological constant can be regarded as a dynamical variable \cite{Henneaux1984,Teitelboim,Henneaux1989} and, in a more recent modern perspective, it is treated as a thermodynamic variable in black hole physics \cite{cald,shuang,Sekiwa,Kastor2009}. In this modern viewpoint, the thermodynamic phase space of black holes is extended by treating the negative cosmological constant as the thermodynamic pressure \cite{Kastor2009,Dolan1,Dolan2,Kubiznak2012}, i.e.,
 \begin{equation}
 \Lambda=-8 \pi P.
 \end{equation}
 By this extension (known as the extended phase space), the mass of AdS black holes should be identified as the enthalpy of spacetime ($M \equiv H$) and the thermodynamic volume (conjugate of the pressure) is defined in a natural way, given by 
 \begin{equation} \label{thermodynamic volume}
 V = {\left( {\frac{{\partial H}}{{\partial P}}} \right)}.
 \end{equation}
 Remarkably, this thermodynamic definition of volume is in agreement with the geometric definition of black hole volume by use of the Komar integral relation (developed by Kastor \cite{Kastor2009}) which is interpreted in terms of the boundary integrals of the renormalized Killing potential at infinity and the Killing potential at the horizon  
 \begin{equation} \label{geometric volume}
 V =  - \left[ {\int_{\partial {\Sigma _\infty }} {d{S_{ab}}\left( {{\omega ^{ab}} - \omega _{{\rm{AdS}}}^{ab}} \right) - \int_{\partial {\Sigma _{\rm{h}}}} {d{S_{ab}}{\omega ^{ab}}} } } \right],
 \end{equation}
 where the Killing potential (${\omega ^{ab}}$) is defined\footnote{The Killing potential was first introduced in Ref. \cite{Bazanski1990}, in which the authors generalized the Komar's conserved quantities to
 spacetimes with a cosmological constant by defining a Gauss-type law for gravity with a cosmological constant. However, in their approach, the cosmological constant is still fixed.} by use of the Killing vector as ${\xi ^a} = {\nabla _b}{\omega ^{ab}}$. \\
In the extended phase space, the first law of black
hole thermodynamics is defined the same as everyday thermodynamics. Consequently, one can think about the other
branches of thermodynamics and apply them in black holes as well. Pushing on this idea, interesting phase transitions
have emerged like Van der Waals phase structure in a number of AdS black hole spacetimes \cite{Kubiznak2012,Kubiznak2017,Gunasekaran,Astefanesei2019}. Another novel direction is studying of black holes as a working substance of a classical heat engine and then evaluating the efficiency for desired cycles \cite{Johnson2014,Johnson2016a,Johnson2016b,HE2019_3Dim,HE2017Hennigar,HE2018,HE2021SZ}.
 \par
 Among all the thermodynamic phenomena applied to anti-de Sitter (AdS) black holes, the Joule-Thomson expansion (also known as the throttling process) has recently gained a considerable interest. In this isenthalpic irreversible expansion the interacting system may cool or heat as it flows from the high pressure to the low pressure region. The first study of Joule-Thomson expansion for AdS black holes was carried out by \"{O}kc\"{u} and Aydıner \cite{JT2017a}. They obtained the inversion temperatures and curves and also the isenthalpic curves for the Reissner-Nordstr\"om AdS black holes and determined the cooling-heating regions. Afterwards, the Joule-Thomson expansion was further examined for Kerr-AdS black holes \cite{JT2018a}, five-dimensional Einstein-Maxwell-Gauss-Bonnet AdS black holes \cite{JT2018c}, AdS black holes with a global monopole \cite{JT2018h}, Quintessence Reissner-Nordstr\"om AdS black holes \cite{JT2018i}, RN-AdS Black Holes in $f(R)$ gravity \cite{JT2018d},  $d$-dimensional charged AdS black holes \cite{JT2018b},  charged Gauss-Bonnet
 black holes in AdS space \cite{JT2018g}, 
 Kerr- Newman-AdS black holes \cite{JT2018e}, Bardeen-AdS black holes \cite{JT2020g}, neutral AdS black holes in massive gravity \cite{JT2019a} and also for a number of other black holes in four- and higher-dimensional spacetimes \cite{JT2018f,	JT2019b,JT2019c,JT2020a,JT2020b,JT2020c,JT2020d,JT2020e,JT2020f,JT2020h,JT2020i,JT2020j,JT2020k,JT2021,arxiv1,arxiv2,arxiv3,arxiv4,arxiv5,arxiv6,arxiv7,arxiv2021}.
 \par
In the present study, we investigate the Joule-Thomson expansion for all cases of BTZ black holes which are (non-) rotating solutions of Einstein (-Maxwell) field equations in $2+1$ spacetime dimensions with a negative cosmological constant \cite{BTZ1992,BHTZ1993,MTZ2000}. In contrast to higher dimensional AdS black hole systems in which the Van der Waals phase transition takes place (except in the Schwarzschild black holes), BTZ black holes reveal no critical behavior. This feature is predictable for the static (non-rotating and neutral) BTZ black holes, since their equation of state is similar to the case of an ideal gas and therefore they are associated with microstructures that do not interact \cite{Ghosh}. For BTZ black holes with electric charge and/or angular momentum, the equation of state
seems like that of the Van der Waals fluid, with the difference that  repulsive interactions are now involved and as a result no phase transition occurs in them.  However, one expects that the Joule-Thomson expansion arises in BTZ black holes with electric charge and/or angular momentum since their underlying microstructures are interacting.  For these reasons, we speculate that the thermodynamic process of Joule-Thomson expansion can in principle take place in BTZ black holes and this could be a nice study to probe the nature of BTZ black holes from a purely thermodynamic argument.
\par
  So far the investigations for holographic Joule-Thomson expansion have concentrated on the cases of 4- and higher-dimensional spacetimes and their lower-dimensional (BTZ) cousins has been considerably less explored. Recently, in Ref. \cite{Liang}, the authors have studied this thermodynamic process only for the case of rotating BTZ black holes.\footnote{It should be noted that while we were preparing this article, we noticed that a study of Joule-Thomson expansion for the case of rotating BTZ black holes had appeared in \underline{arXiv:2104.08841} \cite{Liang}. Our work constitutes a substantial extension of \cite{Liang} and the overlapping results in Sec. \ref{JT in rotating BTZ} were developed independently around the same time. However, here for the first time, the study of Joule-Thomson expansion for the charged BTZ black holes in two different thermodynamic schemes is presented in Sec. \ref{JT in charged BTZ}. Furthermore, here we probe the Joule-Thomson expansion for the charged rotating BTZ black holes in Sec. \ref{JT in charged rotating BTZ}.} But, in this paper we carry out a more extensive study of Joule-Thomson expansion in lower-dimensional black hole spacetimes by considering different backgrounds including rotating, charge and charged rotating BTZ black holes. As will be seen, the case of charged BTZ black holes is more challenging to address due to the fact that these black objects can be studied via two possible thermodynamic schemes \cite{Frassino2015}. As will be evident, the study of this process in them is also more challenging to analyze and this feature makes it difficult to study the BTZ black holes with both charge and rotation parameters. So, we will examine them step by step from the simplest to the most complex. Hence, we will first study the Joule-Thomson expansion for rotating BTZ black holes. Then, the Joule-Thomson expansion will be investigated for the charged (non-rotating) BTZ black holes within two different schemes and it is shown that the results depend definitely on the scheme one is dealing with. We will argue that for the charged BTZ black holes within scheme I, which are thermally unstable, the Joule-Thomson expansion can not occur despite the fact that they are considered as interacting systems. However, we will prove that the Joule-Thomson expansion is present within scheme II, in which thermodynamical instabilities do not arise. To this end, BTZ black hole metrics with both charge and rotation parameters will be examined, too.

\par
Our paper is organized as follows. After briefly reviewing the three-dimensional BTZ spacetimes in Sec. \ref{br}, we shall explore the Joule-Thomson expansion in rotating BTZ black holes in Sec. \ref{JT in rotating BTZ} and investigate the Joule-Thomson coefficient, the inversion curves in the $T-P$ plane and also the isenthalps. In Sec. \ref{JT in charged BTZ}, we will study the Joule-Thomson expansion for charged BTZ black holes within scheme II and will obtain the analytic relations for Joule-Thomson coefficient, inversion curves and the isenthalps. Sec. \ref{JT in charged rotating BTZ} provides discussion on the Joule-Thomson expansion for charged rotating BTZ black holes. Finally, Sec. \ref{clore} gives a summary of the results and the conclusions.

\section{Black holes in $3$-dimensional spacetimes: A brief review}\label{br}
In this section, we shall briefly review the black hole thermodynamics of the three-dimensional BTZ spacetimes. We first study the rotating (electrically neutral) BTZ spacetime which has a straightforward thermodynamic description. After that, we turn our attention to charged (non-rotating) BTZ metrics which have some problematic asymptotic behaviors. Finally, we shall consider BTZ metrics with both the charge and rotation parameters.

\subsection{Rotating BTZ black holes}\label{RBH}

The Einstein-Hilbert action in the three-dimensional AdS background may be written as 
\begin{equation}
{I_{{\rm{EH - AdS}}}} = \frac{1}{{16\pi }}\int {{d^3}x\sqrt { - g} \left[ {R - 2\Lambda } \right]}.
\end{equation}
The resulting field equations, ${G_{\mu \nu }}+ {g_{\mu \nu }} \Lambda = 0$, admit the rotating BTZ spacetime, given by \cite{BTZ1992}
\begin{equation}\label{metricrot}
d{s^2} =  - f(r)d{t^2} + f(r)d{r^2} + {r^2}{\left( { - \frac{J}{{2r ^2}}dt + d\varphi } \right)^2},
\end{equation}
where
\begin{equation}
f(r) =  - 8M + \frac{{{r^2}}}{{{L^2}}} + \frac{{{J^2}}}{{4{r^2}}}.
\end{equation}
The event horizon is located at $r=r_+$, satisfying $f(r_+)=0$. The extended phase space thermodynamics of rotating BTZ black holes have extensively been studied in Ref. \cite{Frassino2015}. Assuming the extended phase space, the enthalpy ($H$), the pressure ($P$), and the thermodynamic volume ($V$) are obtained as
\begin{equation}\label{HPV}
H \equiv M = \frac{{r_ + ^2}}{{8{L^2}}} + \frac{{{J^2}}}{{32r_ + ^2}} \, , \quad P = \frac{1}{{8\pi {L^2}}} \, , \quad V = {\left( {\frac{{\partial H}}{{\partial P}}} \right)_{S,J}} = \pi r_ + ^2.
\end{equation}
The rest of thermodynamic quantities read 
\begin{equation} \label{STO}
S = \frac{\pi }{2}{r_ + } \, , \quad T = {\left( {\frac{{\partial H}}{{\partial S}}} \right)_{P,J}} = \frac{{{r_ + }}}{{2\pi {L^2}}} - \frac{{{J^2}}}{{8\pi r_ + ^3}} \, , \quad \Omega  = {\left( {\frac{{\partial H}}{{\partial J}}} \right)_{S,P}} = \frac{{{J}}}{{16r_ + ^2}}.
\end{equation}
These quantities satisfy the first law of thermodynamics in the extended phase space, i.e.,
\begin{equation}\label{dm}
dM = TdS + VdP + \Omega dJ.
\end{equation}
By combining the above thermodynamic quantities or using the method of scaling argument \cite{Kastor2009}, it can be checked that the corresponding Smarr relation takes the following form
\begin{equation}\label{btzr}
2PV = TS + \Omega J. 
\end{equation}
Obviously, these black hole solutions do not exhibit thermodynamic phase transitions. But, in comparison with the equation of state of an ideal gas (i.e., $PV=NKT$) which is a non-interacting system, it seems rotating BTZ systems are indeed a kind of interacting system. Intuitively, if they are interacting systems, the occurrence of Joule-Thomson expansion must be allowed in them. This is indeed the case and we shall postpone a proof of this to Sec. \ref{JT in rotating BTZ}.

\subsection{Charged BTZ black holes}\label{CBH}

The Einstein-Hilbert action minimally coupled to the Lagrangian of Maxwell's electromagnetism (${\cal F} = F_{\mu \nu}F^{\mu \nu}$) in AdS background is written as
\begin{equation}\label{actioncharge}
{I_{{\rm{EHM - AdS}}}} = \frac{1}{{16\pi }}\int {{d^3}x\sqrt { - g} \left[ {R - 2\Lambda  - \cal F} \right]}.
\end{equation}
The field equations of motion, ${G_{\mu \nu }}+ {g_{\mu \nu }} \Lambda = {T_{\mu \nu }}$, admit the charged (static) BTZ metric as \cite{MTZ2000} 
\begin{equation} \label{charged BTZ metric}
d{s^2} =  - f(r)d{t^2} + f(r)^{-1}d{r^2} + {r^2}d{\varphi ^2},
\end{equation}
where the emblackening factor is 
\begin{equation}
f(r) =  - 8M  + \frac{{{r^2}}}{{{L^2}}}- \frac{{{Q^2}}}{2}\ln \left( {\frac{r}{R_0}} \right),
\end{equation}
in which $R_0$ is an integration constant with units of \textit{length} and it is necessary in order for evaluating the renormalized mass \cite{MTZ2000,Frassino2015,Setare2008}. The one-form gauge potential, which holds in $F=dA$, is given by

\begin{equation} \label{gauge potential}
A =  - \frac{Q}{8}\ln \left( {\frac{r}{{{R_0}}}} \right)dt.
\end{equation}

The renormalized mass of spacetime was calculated before in Refs. \cite{MTZ2000,Frassino2015,Setare2008} as
\begin{equation} \label{mass-charged BTZ}
M = \frac{{{r_{+}^2}}}{{8{L^2}}} - \frac{{{Q^2}}}{{16}}\ln \left( {\frac{r_+}{R_0}} \right).
\end{equation}
The Hawking temperature, the entropy and the electric potential are obtained using the standard methods in general relativity as 
\begin{equation} \label{rest}
T = \frac{{{r_ + }}}{{2\pi {L^2}}} - \frac{{{Q^2}}}{{8\pi {r_ + }}} \, ,\quad S = \frac{\pi }{2}{r_ + } \, , \quad \Phi  =  - \frac{Q}{8}\ln \left( {\frac{{{r_ + }}}{R_0}} \right).
\end{equation}

The role of the integration parameter $R_0$ in the thermodynamic description of electrically charged BTZ black holes is really questionable, especially in the extended phase space. Traditionally, a common choice is setting $R_0=L$, but the electric potential (\ref{gauge potential}) diverges at infinity which causes the electric potential difference as a physical quantity to be ill-defined. The other possible choice is treating $R_0$ as an independent parameter which encloses the charged BTZ system in a circle of radius $R_0$ (actually, this has already been done in order to find the renormalized mass in Refs. \cite{MTZ2000,Setare2008}). These two options have already been discussed in the context of the extended phase space \cite{Frassino2015,HE2017Hennigar,Johnson2020MPLA,Gregory2020} and we concisely review them here.\\

\textbf{Charged BTZ scheme I ($R_0=L$)} \\

In scheme I, the mass-renormalization scale ($R_0$) is identified with the AdS radius $L$. So, from Eq. \ref{mass-charged BTZ}, the enthalpy is straightforwardly obtained in the extended phase space as
\begin{equation}
H(S,P,Q) = \frac{{4P{S^2}}}{\pi } - \frac{{{Q^2}}}{{32}}\ln \left( {\frac{{32P{S^2}}}{\pi }} \right).
\end{equation}
The rest of thermodynamic quantities are simply deduced as
\begin{equation} \label{volume BTZ}
P = \frac{1}{{8\pi L^2}} \, , \quad V = {\left( {\frac{{\partial H}}{{\partial P}}} \right)_{S,Q}} = \frac{{4{S^2}}}{\pi } - \frac{{{Q^2}}}{{32P}},
\end{equation}
and
\begin{equation} \label{temperature BTZ}
S = \frac{\pi }{2}{r_ + } \, ,\quad T= {\left( {\frac{{\partial H}}{{\partial S}}} \right)_{P,Q}} = \frac{{8PS}}{\pi } - \frac{{{Q^2}}}{{16S}} \, , \quad \Phi  = {\left( {\frac{{\partial H}}{{\partial Q}}} \right)_{S,P}} =  - \frac{Q}{{16}}\ln \left( {\frac{{32{S^2}P}}{\pi }} \right).
\end{equation}
There are two serious problems in this scheme. Firstly, the gauge potential (\ref{gauge potential}) diverges at infinity (${r} \to \infty$), so it is not gauge invariant. Secondly, the specific heat at constant volume ($C_V$) is always negative \cite{Johnson2020MPLA}, meaning that charged BTZ black holes within this scheme (I) are thermally unstable (see the subsequent problematic issues in Refs. \cite{Johnson2020MPLA,Gregory2020}). 

It is easy to show that the first law takes the following form
\begin{equation} \label{Smarr BTZ I}
dM = TdS + VdP + \Phi dQ.
\end{equation}
A very simple form for the equation of state, or equivalently, for the Smarr formula, is found as
\begin{equation}\label{PVTS}
2PV= TS.
\end{equation}
Despite the complicated form of thermodynamic volume (\ref{volume BTZ}) in this scheme, this is an interesting result since the Smarr formula takes a very simple form which reminds us that of non-interacting ideal gas system. This means that, most likely, the Joule-Thomson expansion cannot be observed in this scheme, as will be shown in Sec. \ref{JT in charged BTZ}. This is an inconsistent result in this scheme since, according to the previous studies \cite{Ghosh}, we already knew that charged BTZ black holes are indeed interacting systems. So, we conclude that this scheme (I) is not correct for thermodynamical considerations of charged BTZ black holes. \\

\textbf{Charged BTZ scheme II  (independent $R_0$)}\\

In scheme II, the renormalization length scale ($R_0$) is identified as a new thermodynamic parameter \cite{Frassino2015}, which leads to retention of the standard definition of the thermodynamic volume in accordance to the geometric formula (\ref{geometric volume}), yielding $V=\pi r_+^2$. Assuming this scheme, from Eq. \ref{mass-charged BTZ}, the enthalpy is straightforwardly obtained in the extended phase space as
\begin{equation}\label{MCBH}
H(S,P,Q,{R_0}) = \frac{{4P{S^2}}}{\pi } - \frac{{{Q^2}}}{{16}}\ln \left( {\frac{{2S}}{{\pi {R_0}}}} \right).
\end{equation}
The thermodynamic quantities read
\begin{equation}\label{PVCBH}
P = \frac{1}{{8\pi {L^2}}} \, , \quad V = {\left( {\frac{{\partial H}}{{\partial P}}} \right)_{S,Q,{R_0}}} = \pi r_ + ^2 = \frac{{4{S^2}}}{\pi } \, ,
\end{equation}
and
\begin{equation}\label{KPhi}
K = {\left( {\frac{{\partial H}}{{\partial {R_0}}}} \right)_{S,Q,P}} = \frac{{{Q^2}}}{{16{R_0}}} \, , \quad \Phi  = {\left( {\frac{{\partial H}}{{\partial Q}}} \right)_{S,P,{R_0}}} =  - \frac{Q}{8}\ln \left( {\frac{{2S}}{{\pi {R_0}}}} \right),
\end{equation}
where $K$ is the thermodynamic conjugate to $R_0$. The temperature and the entropy are the same as before in Eq. (\ref{rest}). Therefore, the charged BTZ black hole system in this scheme has four pairs of intensive-extensive variables, i.e., $T-S$, $P-V$, $\Phi-Q$ and the new pair $K-R_0$. In thermodynamics, a fundamental equation containing all the thermodynamic information relates all extensive properties of a physical system. Hence, the fundamental equation in terms of energy in scheme II may be written as
\begin{equation}
E = H - PV =  - \frac{{{Q^2}}}{{16}}\ln \left( {\frac{{2S}}{{\pi {R_0}}}} \right).
\end{equation}
Note that since $\left( {\frac{{2S}}{{\pi {R_0}}}} \right) = \frac{{{r_ + }}}{{{R_0}}}$ and $R_0 > r_+$, so the internal energy is always positive.

Now, the first law of thermodynamics is verified as
\begin{equation}\label{dcbh}
dM = TdS + VdP + \Phi dQ + Kd{R_0},
\end{equation}
and the Smarr formula takes the following formula
\begin{equation}\label{scbh}
2PV = TS + K{R_0},
\end{equation}
in agreement with the method of scaling argument \cite{Kastor2009}.
Evidently, in this scheme, the thermodynamic volume and the entropy are not independent variables (a familiar result, typical of static black hole spacetimes), and so ${C_V}=0$. As a result, there is not any fundamental thermodynamic instability in this scheme in contrast to scheme I in which thermal instabilities always are present. On the other hand, since a circle of radius $R_0$ encloses the charged BTZ system, the potential (\ref{gauge potential}) at infinity (i.e., at the boundary $r=R_0$) vanishes, which makes it possible to define a gauge invariant thermodynamic potential, as seen in Eq. (\ref{rest}).\\

\textbf{Complementary discussion on charged BTZ schemes} \\

One can show that, besides the specific heat, all the thermodynamic coefficients are physical in the second scheme (II) but some of them are unphysical in the first scheme (I). For example, in scheme I, it turns out that the specific heat at constant volume is negative ($C_V<0$) but $C_P>0$ \cite{Johnson2020MPLA}. In addition, explicit computation shows that the adiabatic compressibility is always negative in scheme I, i.e.,
\begin{equation}
{\kappa _S} =  - \frac{1}{V}{\left( {\frac{{\partial V}}{{\partial P}}} \right)_S} =  - \frac{{{Q^2}}}{{32V{P^2}}} < 0.
\end{equation}
So, the system in scheme I is always unstable. But, since the entropy and the thermodynamic volume are not independent in scheme II (see Eq. (\ref{PVCBH})), the adiabatic compressibility and the heat capacity at constant volume vanish, ${C_V} = {\kappa _s} = 0$ (this property is typical of static black holes in 4- and higher dimensions.) In this scheme (II), one can show that there always exist physical black hole regions in which the heat capacity at constant pressure, the isobaric expansivity and the isothermal compressibility are positive. In addition, entropy is a smooth, concave function of extensive variables in scheme II. These are the essential requirements to guarantee the local thermal stability of a canonical system, which here are valid for charged BTZ black holes in scheme II. However, for a full discussion, the determinant of the Hessian matrix of the entropy (or energy) must be examined explicitly. For these reasons, scheme II that is free from pathological behaviors should be considered as the more logical scheme.

Furthermore, working in scheme II has some interesting consequences. The renormalized mass (\ref{mass-charged BTZ}) by introducing the parameter $R_0$ refers to the fact that the total gravitational energy together with the electromagnetic energy of spacetime are enclosed in a circle of radius $R_0$. Therefore, a straightforward interpretation comes to mind for the changes of this parameter (specific to scheme II): by changing the energy content of black hole spacetime according to the first law of thermodynamics, the radius of the circle enclosing the total energy also changes. On the other hand, from the comparison of the Smarr relation (\ref{scbh}) (which follows from scaling argument \cite{Kastor2009}) and the first law (\ref{dcbh}), in order for these two relations to be compatible with each other, the parameter $R_0$ must be considered as a thermodynamic variable which clearly implies that the second scheme is the self-consistent one.

We shall explore the thermodynamic process of Joule-Thomson expansion in charged BTZ black holes within this scheme (II) in Sec. \ref{JT in charged BTZ} which leads to a consistent result. However, in scheme I, one obtains inconsistent results, as will be discussed further in the sections \ref{JT in charged BTZ} and \ref{clore}.

\subsection{Charged rotating BTZ black holes}
The (2+1)-dimensional black hole solutions with both charge and rotation parameters (found by Martinez, Teitelboim, and Zanelli in \cite{MTZ2000}) are described by the bulk action (\ref{actioncharge}) and the line element is the same as (\ref{metricrot}) with
\begin{equation}
f(r) =  - 8M + \frac{{{r^2}}}{{{L^2}}} + \frac{{{J^2}}}{{4{r^2}}} - \frac{{{Q^2}}}{2}\ln \left( {\frac{r}{{{R_0}}}} \right).
\end{equation}
The enthalpy and the Hawking temperature of the charged rotating BTZ black hole are given by
\begin{equation}\label{MTcr}
M = \frac{{{r_ + }^2}}{{8{L^2}}} + \frac{{{J^2}}}{{32{r_ + }^2}} - \frac{{{Q^2}}}{{16}}\ln \left( {\frac{{{r_ + }}}{{{R_0}}}} \right), \quad T = \frac{{{r_ + }}}{{2\pi {L^2}}} - \frac{{{J^2}}}{{8\pi {r_ + }^3}} - \frac{{{Q^2}}}{{8\pi {r_ + }}}.
\end{equation}
With the pressure and the volume given by (\ref{PVCBH}), the intensive parameter $K$ and the electric potential obtained in (\ref{KPhi}) and the angular velocity of (\ref{STO}), the first law of thermodynamics
\begin{equation}
dH = TdS + VdP + \Phi dQ + \Omega dJ + Kd{R_0},
\end{equation}
and the Smarr relation 
\begin{equation}
2PV = TS + \Omega J + K{R_0},
\end{equation}
are verified.
\section{Joule-Thomson expansion in rotating BTZ black holes}\label{JT in rotating BTZ}
In the Joule-Thomson expansion the temperature of the gas changes as the gas flows through a throttle valve from a high pressure region to the low pressure one with larger volume. During this irreversible adiabatic expansion, the gas does work and the enthalpy $H=U+PV$ remains constant. For an ideal gas this expansion does not yield any change in temperature. This can be checked by the following thermodynamic identity
\begin{equation}\label{dh}
dH=C_p dT+V(1-\alpha T)dp,
\end{equation}
where $C_p$ is the heat capacity at constant pressure and $\alpha=1/V (\partial V/\partial T)_P$ is the thermal expansion coefficient. For the ideal gas $\alpha T=1$ and therefore $dH=C_p dT$, i.e., if the enthalpy is unchanged, so is the temperature. However, for the real gases the second term of Eq. \ref{dh} does not vanish which means that the Joule-Thomson expansion will result in cooling/heating as a consequence of interactions between gas molecules. 
\par 
The Joule-Thomson coefficient measures the change in the gas temperature while reducing pressure at constant enthalpy and is defined as
\begin{equation}
\mu=\Big(\frac{\partial T}{\partial P}\Big)_H.
\end{equation} 
 This coefficient is positive (negative) when cooling (heating) occurs as the pressure decreases. Writing $\mu$ in terms of thermodynamic variables such as, e.g., pressure, volume, charge, angular momentum and other related quantities, and equating to zero, one finds the inversion temperature which depends on the pressure. The curve on which $\mu$ vanishes in the $T-P$ plane is called the inversion curve. In what follows we study the Joule-Thomson expansion in rotating, charged and charged rotating BTZ black holes.
 \par 
 Using Eqs. \ref{HPV} and \ref{STO}, the equation of state of rotating BTZ black hole is given by
 \begin{equation}\label{eosr}
 P=\frac{l_p T}{v}+\frac{8J^2 l_p^4}{\pi v^4},
 \end{equation}
 where $v=4 r_+ l_p$ is the specific volume and $l_p$ is the Planck length. Comparing (\ref{eosr}) with the equation of state of a non-ideal fluid, it is evident that the rotating BTZ black hole is an interacting system, as stated in Sec. \ref{RBH}, and therefore the Joule-Thomson expansion can arise. Since the mass of an AdS black hole represents its enthalpy, during the Joule-Thomson expansion the black hole mass remains constant, i.e., $dM=0$. As a result, Eq. (\ref{dm}) reads
 \begin{equation}\label{tds}
 TdS=-VdP,
 \end{equation}
 for $dJ=0$. Differentiating Eq. (\ref{btzr}) besides Eq. (\ref{tds}) gives
 \begin{equation}
 -3V+S\left(\frac{\partial T}{\partial P}\right)_M-2P\left(\frac{\partial V}{\partial P}\right)_M+ J\left(\frac{\partial \Omega}{\partial P}\right)_M=0, 
 \end{equation}
 which leads to the Joule-Thomson coefficient in terms of the rotating BTZ black hole parameters
 \begin{equation}\label{muu}
\mu=\left(\frac{\partial T}{\partial P}\right)_M=\frac{1}{S}\left[2P\left(\frac{\partial V}{\partial P}\right)_M-J\left(\frac{\partial \Omega}{\partial P}\right)_M+3V\right].
\end{equation}
Rewriting $V$ and $\Omega$ in Eqs. (\ref{HPV}) and (\ref{STO}) in terms of the entropy and differentiating, the partial derivatives in Eq. (\ref{muu}) are then given by
\begin{eqnarray}\label{dpdv}
\left(\frac{\partial V}{\partial P}\right)_M&=&\frac{8S}{\pi}\left(\frac{\partial S}{\partial P}\right)_M,\nonumber\\
\left(\frac{\partial \Omega}{\partial P}\right)_M&=&\frac{-2\Omega}{S}\left(\frac{\partial S}{\partial P}\right)_M,
\end{eqnarray}
where 
\begin{equation}\label{dsdp}
\left(\frac{\partial S}{\partial P}\right)_M=\frac{2S^3}{\pi M-8 P S^2},
\end{equation}
is obtained from differentiating $H$ in Eq. (\ref{HPV}). combining Eq. (\ref{dpdv}) with (\ref{dsdp}) and inserting into Eq. (\ref{muu}), the Joule-Thomson coefficient in terms of the rotating BTZ black hole parameters can be rewritten as
\begin{equation}
\mu=\frac{{4S}}{\pi }\left( {1 + \frac{{4{\pi ^3}{J^2}}}{{{\pi ^3}{J^2} - 512P{S^4}}}} \right).
\end{equation} 
Also it is possible to evaluate the Joule-Thomson coefficient from
\begin{equation}\label{musw}
\mu  = {\left( {\frac{{\partial T}}{{\partial {r_ + }}}} \right)_{H,J}}{\left( {\frac{{\partial {r_ + }}}{{\partial P}}} \right)_{H,J}},
\end{equation}
where  $P(M,r_{+})$ is obtained by replacing $L$ in terms of $P$ in the first relation in Eq. (\ref{HPV}) and then solving for $P$ and $T(M,r_{+})$ is determined by replacing $L$ in terms of $P(M,r_{+})$ in the second relation in Eq. (\ref{STO}).

\par
\begin{figure}
	\includegraphics[scale=0.9]{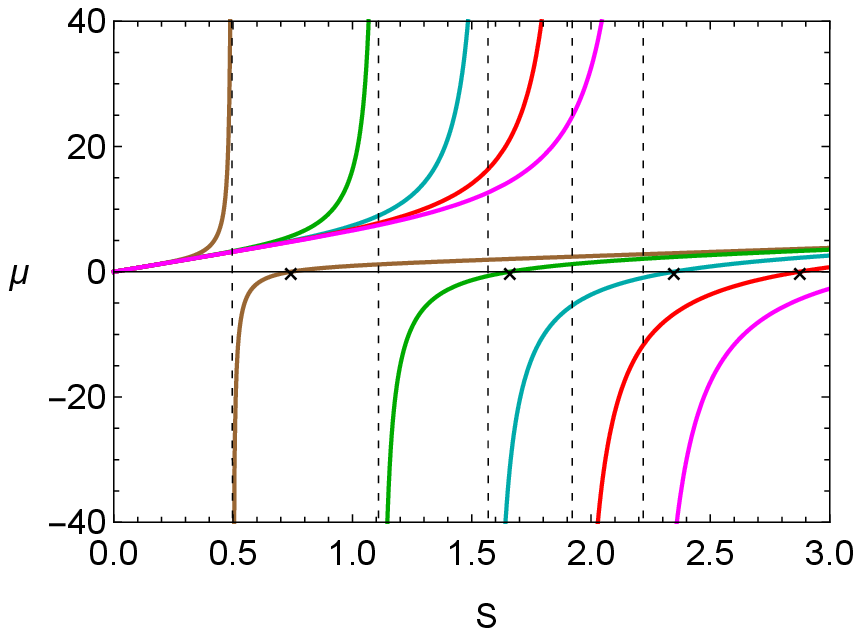}
	\hskip 1 cm
	\includegraphics[scale=0.85]{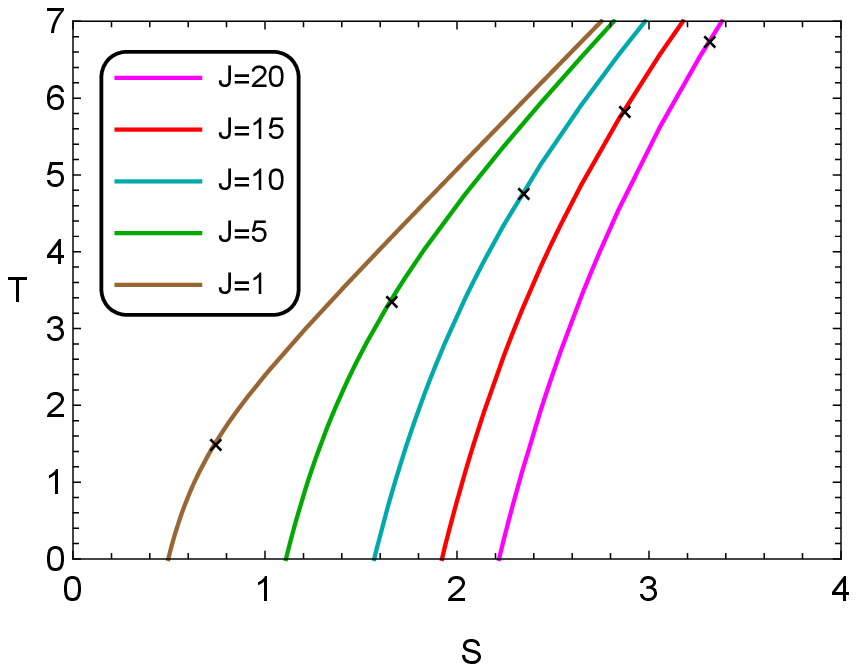}
	\caption{\textbf{$\mu$ vs $S$ (left panel) and $T$ vs $S$ (right panel) for rotating BTZ black holes}: Diagrams are plotted with fixed  $P=1$ and different values of $J$. The divergent points of the JT coefficient in the left panel (vertical dashed lines) are consistent with the zero point of the Hawking temperatures in the right panel. In the physical domain with positive temperature, $\mu$ increases monotonically and becomes zero only once for each value of $J$ (illustrated with cross marks). Therefore there are only minimum inversion temperatures for rotating  BTZ black holes, i.e., the BTZ black holes always cool at large entropies $(\mu >0)$ and heat at small ones $(\mu <0)$ . The corresponding minimum inversion temperatures for different values of $J$ are shown by cross marks in the right panel as well.}
	\label{rot_muTS}
\end{figure}

Fig. \ref{rot_muTS} shows the Joule-Thomson coefficient of the rotating BTZ black hole as a function of entropy for a fixed value of pressure and different values of $J$. Also, the Hawking temperature versus entropy is illustrated in the right panel of this figure. The divergent points of JT coefficient in the left panel (vertical dashed lines) correspond to the points with zero temperature in the right panel. For each $J$, the region to the left of the vertical dashed line is unphysical since the corresponding  temperatures are negative. In the physical domain (to the right of the vertical dashed line), $\mu$ increases and becomes zero at certain entropies for different values of $J$ (illustrated with cross marks). Therefore there are only minimum inversion temperatures for  rotating  BTZ black holes, i.e., the BTZ black holes cool at large entropies $(\mu >0)$.

\begin{figure}
	\includegraphics[scale=0.5]{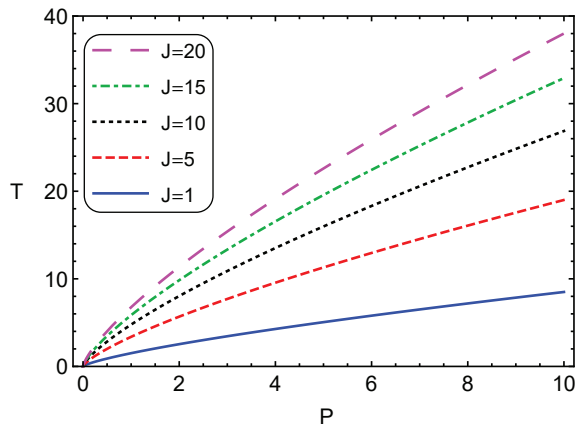}
	\caption{The inversion curves of the rotating BTZ black holes for five different values of $J$. For each specified $J$, cooling and heating occurs above and below the inversion curve, respectively.}
	\label{rot_inv_J}
\end{figure}

\begin{figure}
	\includegraphics[scale=0.37]{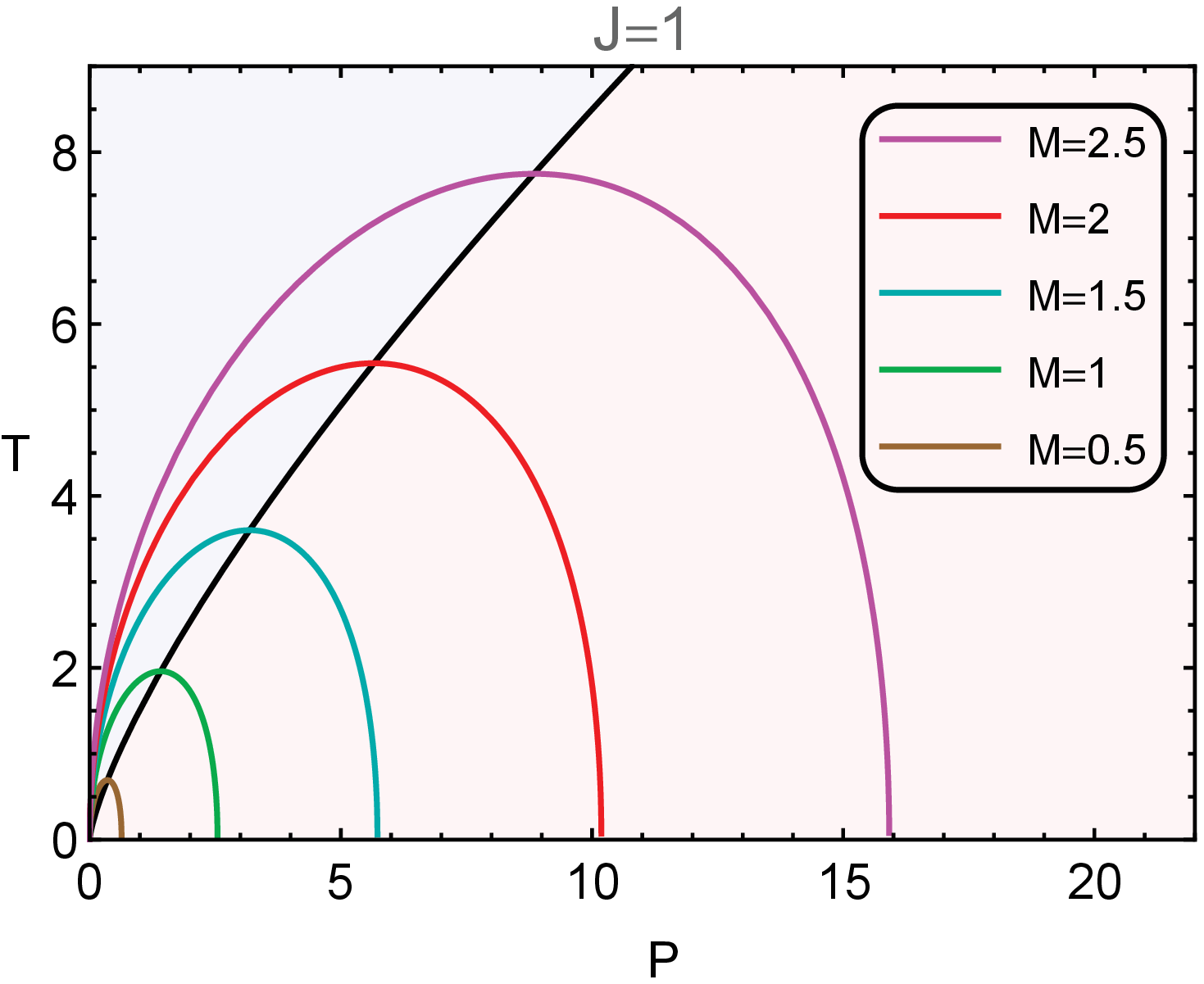}
	\hskip 0.5 cm
	\includegraphics[scale=0.37]{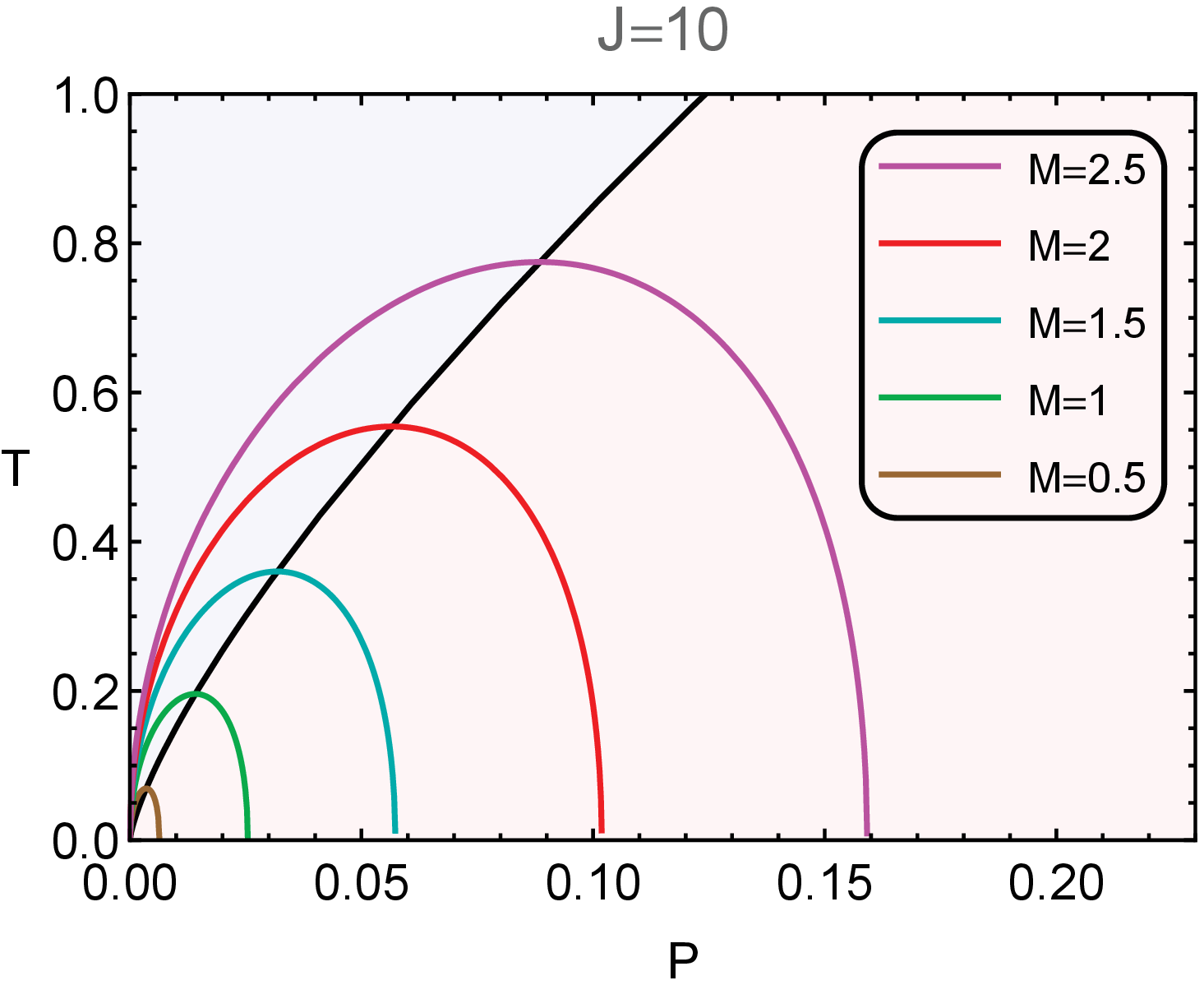}
	\hskip 0.5 cm
	\includegraphics[scale=0.37]{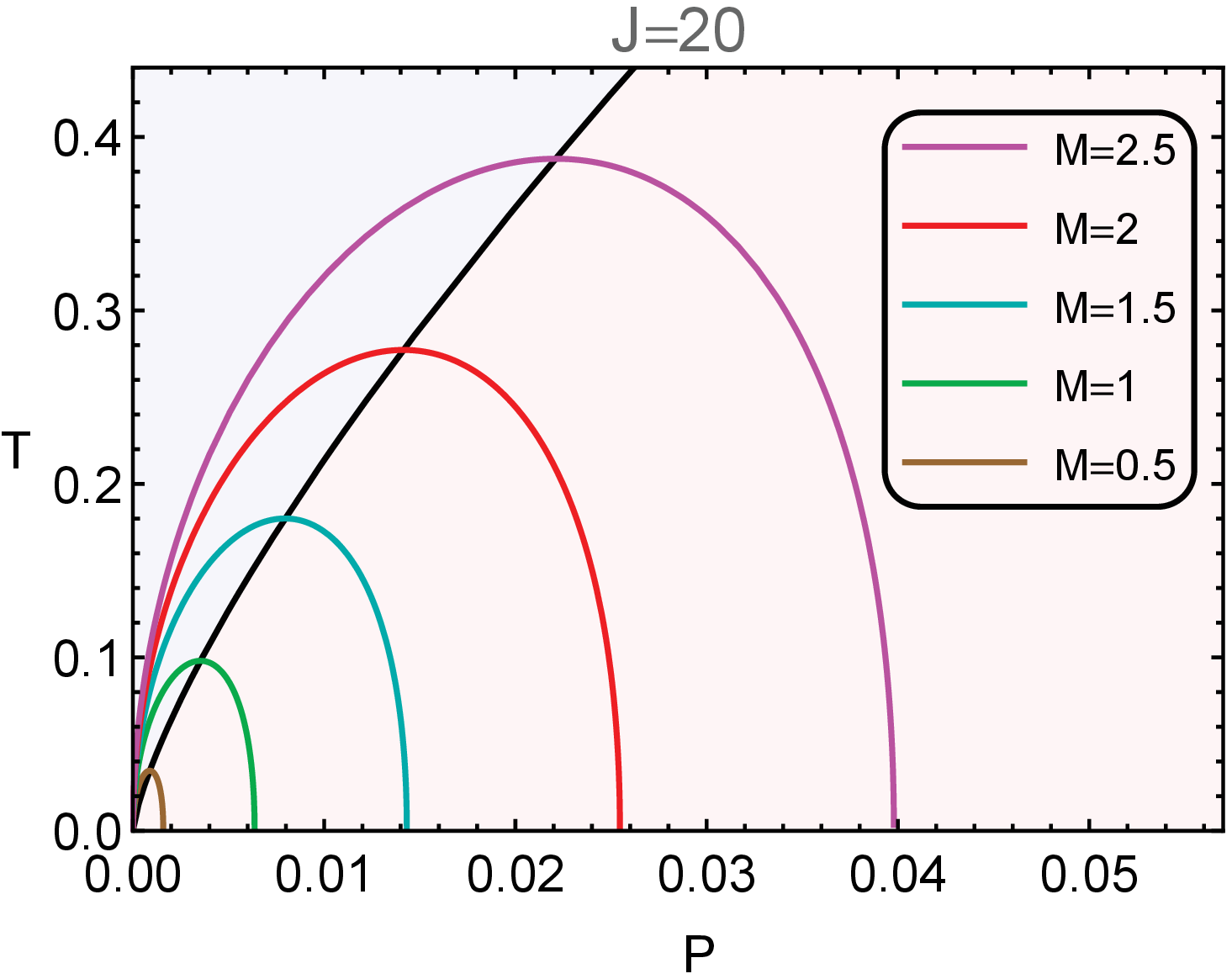}
	\caption{The inversion curves (black curves) and the isenthalps (colored curves) of rotating BTZ black holes with different values of $J$. For the blue regions, decreasing pressure at constant enthalpy leads to cooling ($\mu>0$) and for the red regions heating occurs as a result of reducing pressure while keeping the enthalpy constant ($\mu<0$).}
	\label{rot_isenthalps}
\end{figure}

The invesion pressure $P_i$ is found by letting $\mu=0$ 
\begin{eqnarray}\label{pinv}
P_i&=&\frac{1}{2}\left(\frac{\partial P}{\partial V}\right)_M\left[J\left(\frac{\partial \Omega}{\partial P}\right)_M-3V\right]\nonumber\\
&=&\frac{5 J^2 \pi^3}{512 S^4}
\end{eqnarray}
and, finally, make use of Eq.(\ref{STO}) for $T$, the inversion temperature can be expressed in terms of $P_i$
\begin{equation}\label{ti}
T_i=\frac{{4{{\left( {\frac{2}{5}} \right)}^{3/4}}{J^{1/2}}{P_i}^{3/4}}}{{{\pi ^{1/4}}}}.
\end{equation} 
The inversion curves for different values of $J$ are displayed in Fig. \ref{rot_inv_J}. For each specified $J$, cooling and heating occurs above and below the inversion curve, respectively. Contrary to Van der Waals fluids with closed inversion curves in the $T-P$ plane, here the inversion curves are increasing functions with one branch, i.e, as stated before, there exists minimum inversion temperatures above which the Joule-Thomson expansion leads to cooling. The similar qualitative behavior of the inversion curves of the rotating BTZ black holes with those of charged AdS \cite{JT2017a}, Kerr-AdS \cite{JT2018a}, Kerr-Newman-AdS \cite{JT2018e}, Bardeen-AdS \cite{JT2020g}, Born-Infeld AdS \cite{JT2021} and charged Gauss-Bonnet AdS \cite{JT2018g} black holes is enforced by the cosmological constant. Also as seen in this figure, if pressure decreases, the inversion temperature decreases, too, i.e., cooling occurs for the larger domain. Clearly, from Eq.(\ref{ti}), the minimum inversion temperature $T_i^{\min } = {\left. {{T_i}} \right|_{{p_i} \to 0}}$ for the rotating BTZ black hole is zero and all the inversion curves in Fig. \ref{rot_inv_J} pass through the origin.
Rewriting $M$ in Eq. (\ref{HPV}) in terms of $S$ and $P$ and solving for $S$, gives
\begin{equation}
S = \frac{1}{4}\sqrt {\frac{\pi }{2}} \sqrt {\frac{{4M + \sqrt {16{M^2} - 2\pi{J^2}P } }}{P}},
\end{equation}
which when inserted into the temperature relation $T(S,P)$ in Eq. (\ref{STO}), isenthalpic curves in the $T-P$ plane are obtained via
\begin{equation}
T = 4\sqrt {\frac{2}{\pi }} \frac{{\left( {8{M^2} - \pi {J^2}P + 2M\sqrt {16{M^2} - 2\pi {J^2}P} } \right)}}{{\sqrt P {{\left( {4M + \sqrt {16{M^2} - 2\pi {J^2}P} } \right)}^{3/2}}}}.
\end{equation}
The isenthalpic curves of rotating BTZ black holes for $J=1,10$ and $20$ are shown in Fig. \ref{rot_isenthalps}.  When the isenthalpic curves intersect the inversion curve, their gradients change sign. For the blue regions where the gradients of the isenthalps are positive, decreasing pressure at constant enthalpy leads to cooling ($\mu>0$). For the red regions with negative gradients of the isenthalps, heating occurs as a result of reducing pressure while keeping the enthalpy constant ($\mu<0$).  

\section{Joule-Thomson expansion in charged BTZ black holes}\label{JT in charged BTZ}
In this section we will discuss the Joule-Thomson expansion for the charged BTZ black holes within scheme (II) for which $R_{0}$ is a new thermodynamic parameter. As stated in Sec. \ref{CBH}, there is no Joule-Thomson effect for charged BTZ black holes within scheme (I). The proof of this will be postponed until the end of this section.

The equation of state of charged rotating BTZ black hole, i.e., \footnote{The equation of state in scheme II remains the same as in scheme I. This is because the equation of state is derived from the temperature (\ref{rest}), and the temperature is the same for both schemes.}
\begin{equation}\label{eqosCBH}
P = \frac{{{l_p}T}}{v} + \frac{{{Q^2}{l_p}^2}}{{2\pi {v^2}}},
\end{equation}
reveals its interacting nature and indeed the Joule-Thomson expansion may happen. Similar to the procedure used in Sec. \ref{JT in rotating BTZ}, the Joule-Thomson coefficient for charged BTZ black holes within scheme (II) is obtained by differentiating Eq. (\ref{scbh}),
with $R_{0}$ held constant and replacing $TdS$ by $-VdP$ which is found from Eq. (\ref{dcbh}) for $dM=dQ=dR_{0}=0$. The result reads
\begin{equation}\label{muc1}
\mu=\left(\frac{\partial T}{\partial P}\right)_M=\frac{1}{S}\left[2P\left(\frac{\partial V}{\partial P}\right)_M+3V\right],
\end{equation}
where ${\left( {\partial V/\partial P} \right)_M}$ is the same as in Eq. (\ref{dpdv}) and
\begin{equation}\label{spmc}
{\left( {\frac{{\partial S}}{{\partial P}}} \right)_M} = \frac{{64{S^3}}}{{\pi {Q^2} - 128 P{S^2}}},
\end{equation}
is calculated by differentiating Eq. (\ref{MCBH}), with $H\equiv M$ held constant. Putting (\ref{spmc}) into the first line of (\ref{dpdv}) and then replacing (\ref{dpdv}) in Eq. (\ref{muc1}), yields
\begin{equation}\label{muc}
\mu  = 4S\left( {\frac{1}{\pi } + \frac{{2{Q^2}}}{{\pi {Q^2} - 128P{S^2}}}} \right),
\end{equation}
which also can be checked by using Eq. (\ref{musw}).
\begin{figure}
	\includegraphics[scale=0.9]{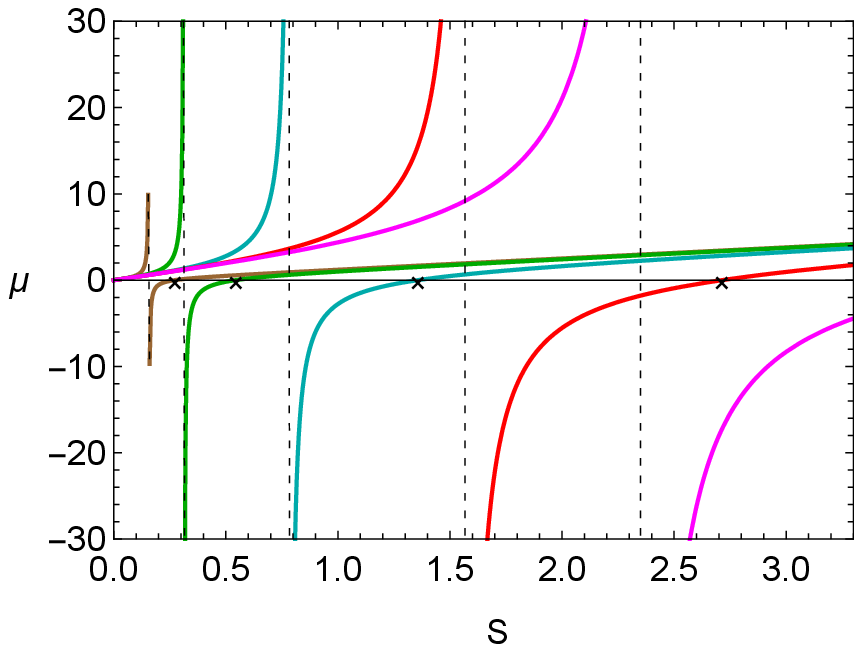}
	\hskip 1 cm
	\includegraphics[scale=0.9]{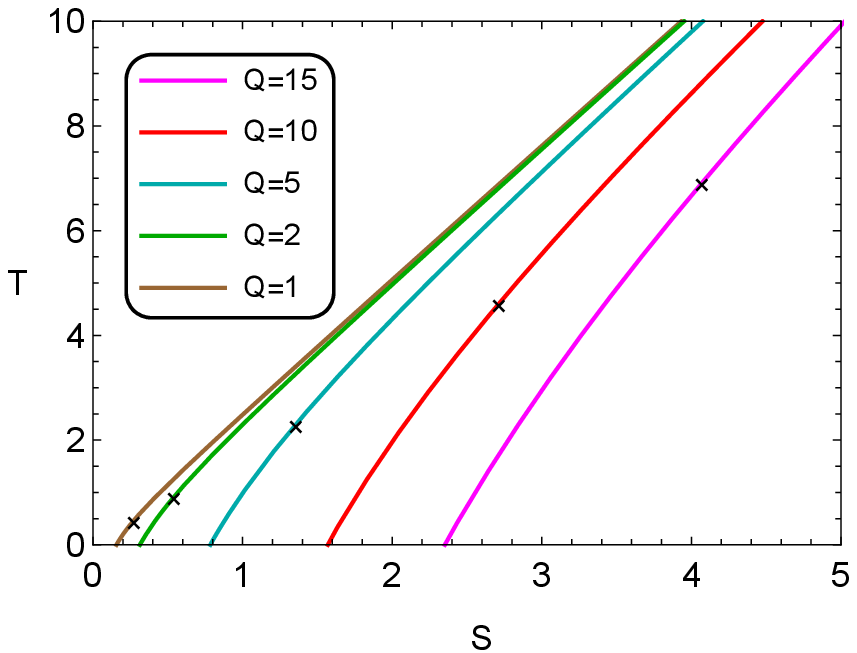}
	\caption{ \textbf{$\mu$ vs $S$ (left panel) and $T$ vs $S$ (right panel) for charged BTZ black holes}: Diagrams are plotted with fixed  $P=1$ and different values of $Q$. The divergent points of the JT coefficient in the left panel (vertical dashed lines) are consistent with the zero point of the Hawking temperatures in the right panel. In the physical domain with positive temperature, $\mu$ increases monotonically and becomes zero only once for each value of $Q$ (illustrated with cross marks). Therefore there are only minimum inversion temperatures for charged BTZ black holes, i.e., the BTZ black holes always cool at large entropies $(\mu >0)$ and heat at small ones $(\mu <0)$ . The corresponding minimum inversion temperatures for different values of $Q$ are shown by cross marks in the right panel as well.}
	\label{char_muTS}
\end{figure}
The Joule-Thomson coefficient versus $S$ is shown in the left panel of Fig. \ref{char_muTS} for different values of $Q$. It is clear that there are only minimum inversion temperatures for charged BTZ black holes, as in the case of rotating BTZ black holes. These minimum inversion temperatures are shown with cross marks in the graph of $T$ versus $S$ in the right panel of Fig. \ref{char_muTS}. We also show the vertical asymptotes of $\mu$  as dashed lines in the left panel which correspond to the points of zero temperatures in the right panel.\\
By setting $\mu$ equal to zero in Eq. (\ref{muc}), the inversion pressure and temperature are obtained
\begin{eqnarray}
{P_i}& =&  - \frac{3}{2}V{\left( {\frac{{\partial P}}{{\partial V}}} \right)_M}=\frac{{3\pi {Q^2}}}{{128{S^2}}},\nonumber\\
{T_i} &=& \sqrt {\frac{2}{{3\pi }}} Q{P_i}^{1/2}.
\end{eqnarray}

\begin{figure}
	\includegraphics[scale=0.5]{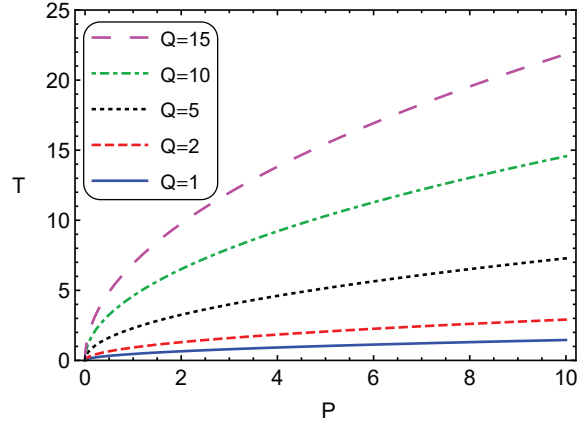}
	\caption{The inversion curves of the charged BTZ black holes for five different values of $Q$. For each specified $Q$, cooling and heating occurs above and below the inversion curve.}
	\label{char_inv_Q}
\end{figure}
\begin{figure}
	\includegraphics[scale=0.37]{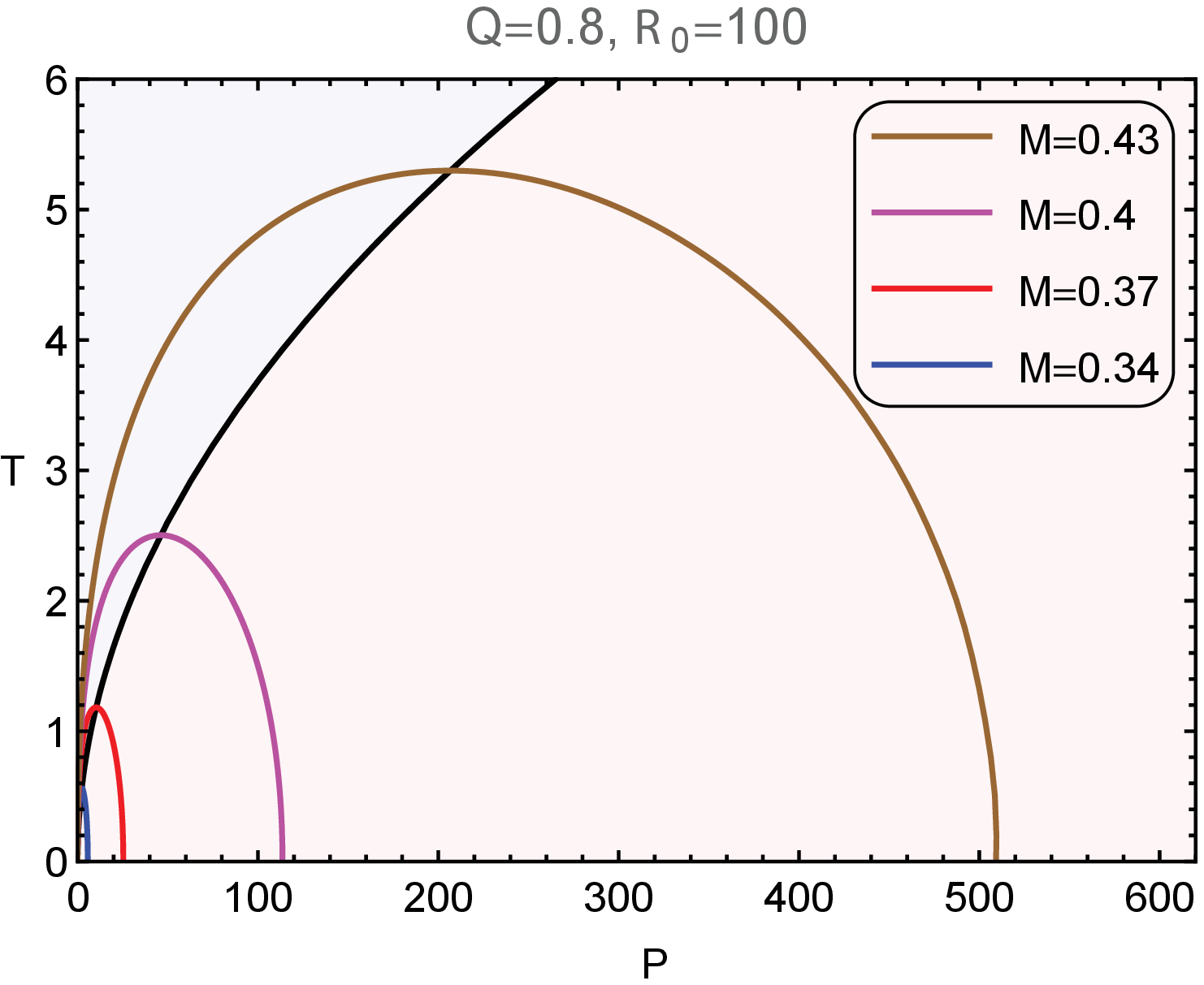}
	\hskip 0.5 cm
	\includegraphics[scale=0.37]{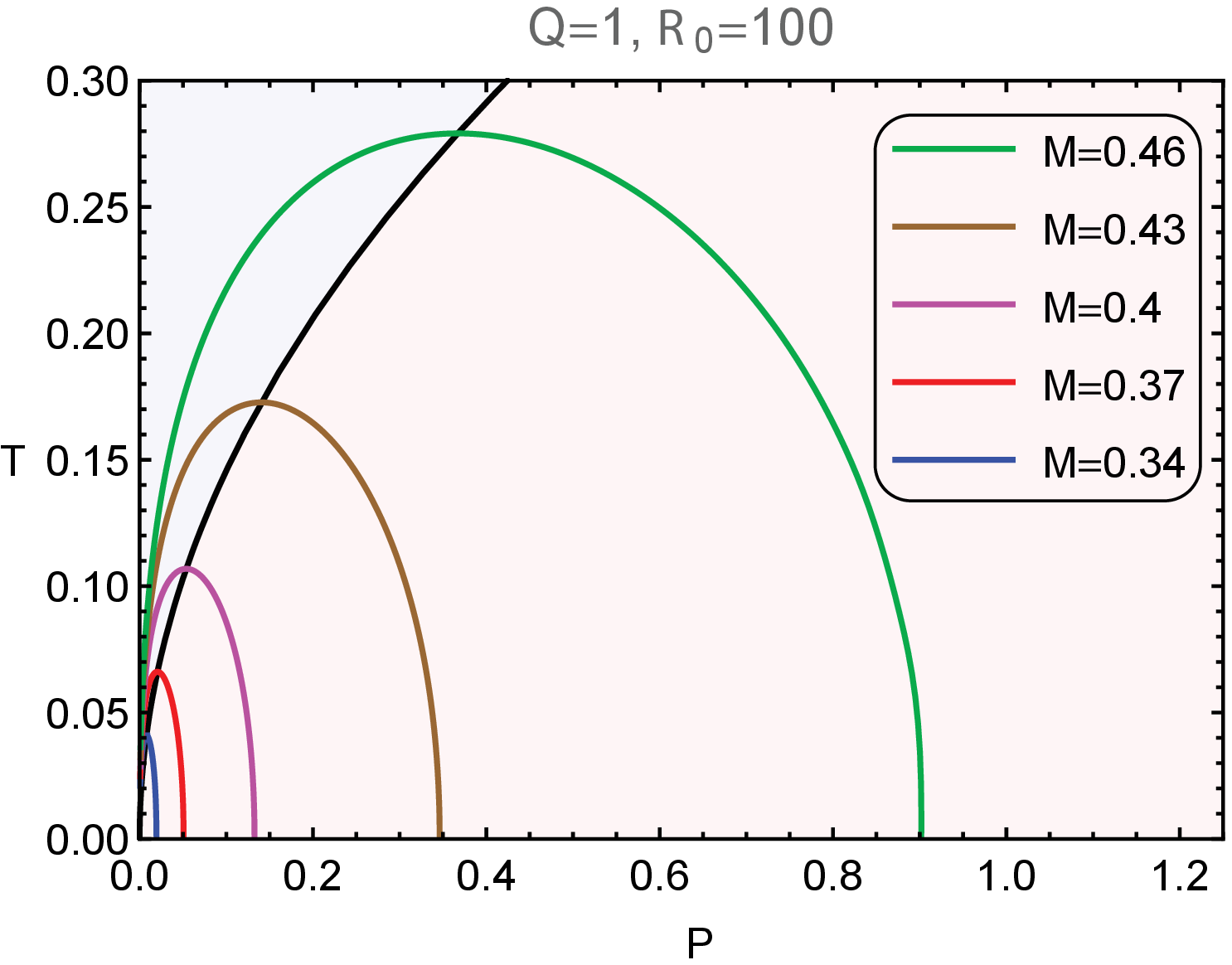}
	\hskip 0.5 cm
	\includegraphics[scale=0.37]{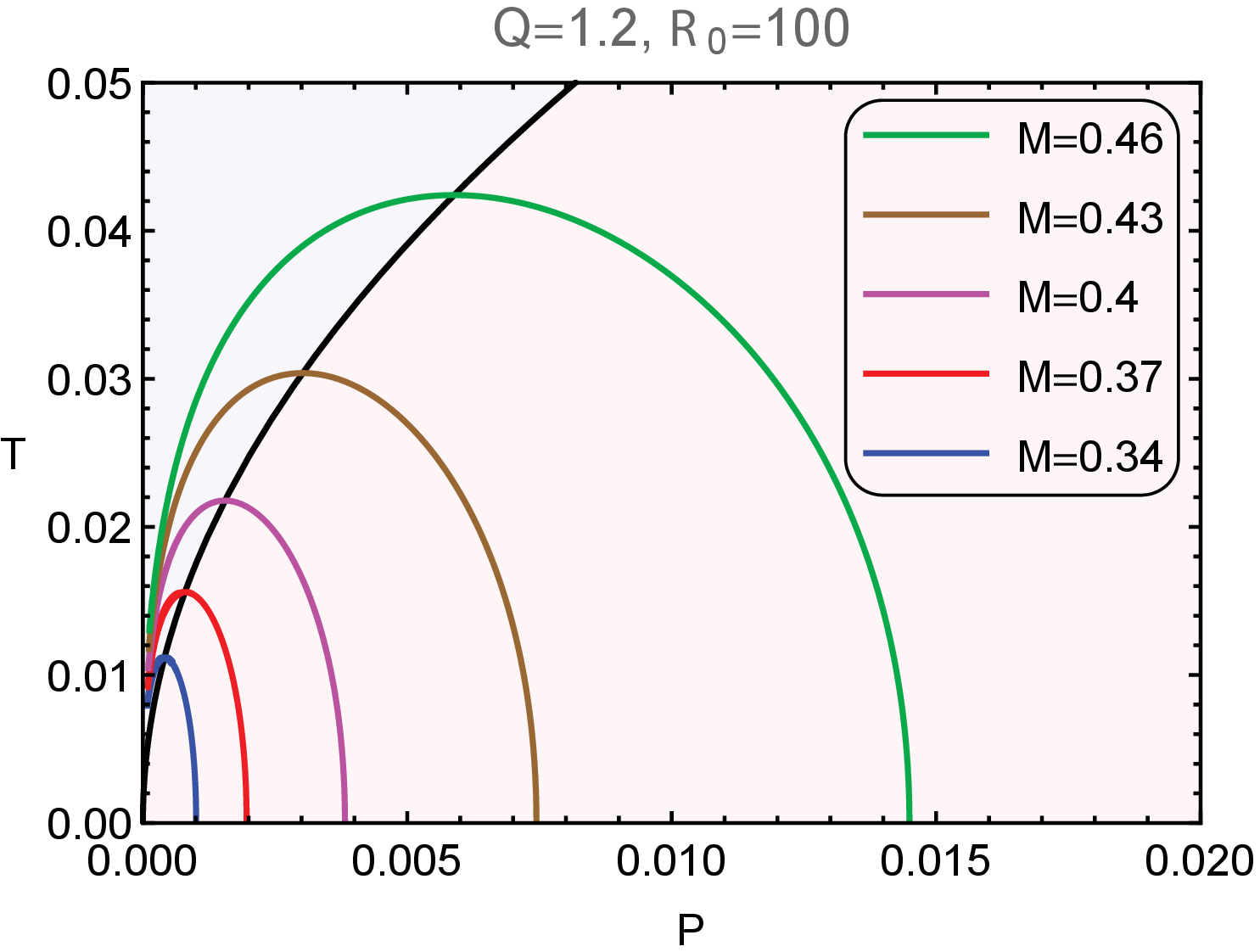}
	\caption{The inversion curves (black curves) and the isenthalps (colored curves) of charged BTZ black holes with $R_0=100$ and different values of $Q$. For the blue regions, decreasing pressure at constant enthalpy leads to cooling ($\mu>0$) and for the red regions heating occurs as a result of reducing pressure while keeping the enthalpy constant ($\mu<0$). }
	\label{char_isenthalps_1}
\end{figure}
\begin{figure}
	\includegraphics[scale=0.37]{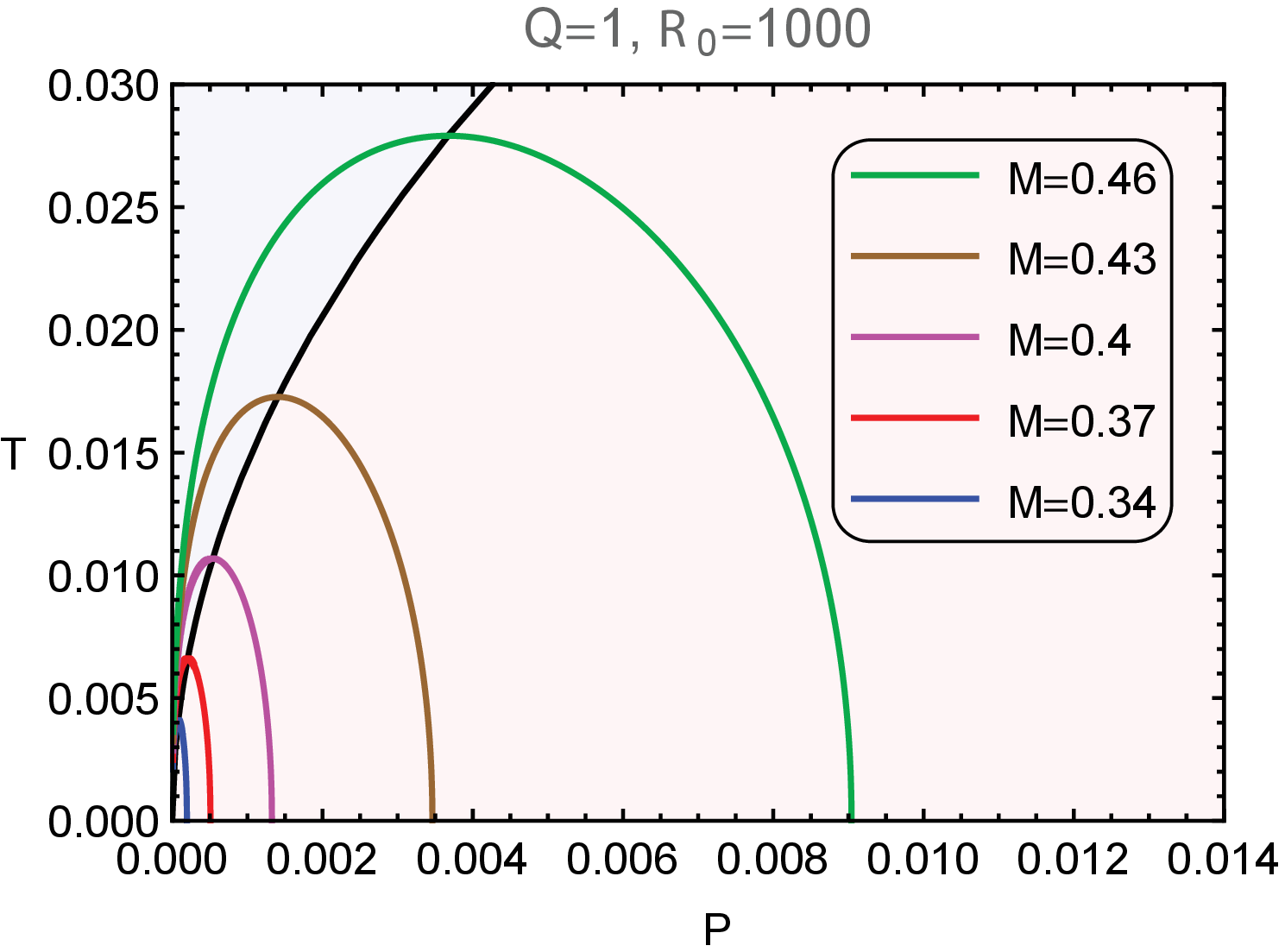}
	\hskip 0.5 cm
	\includegraphics[scale=0.35]{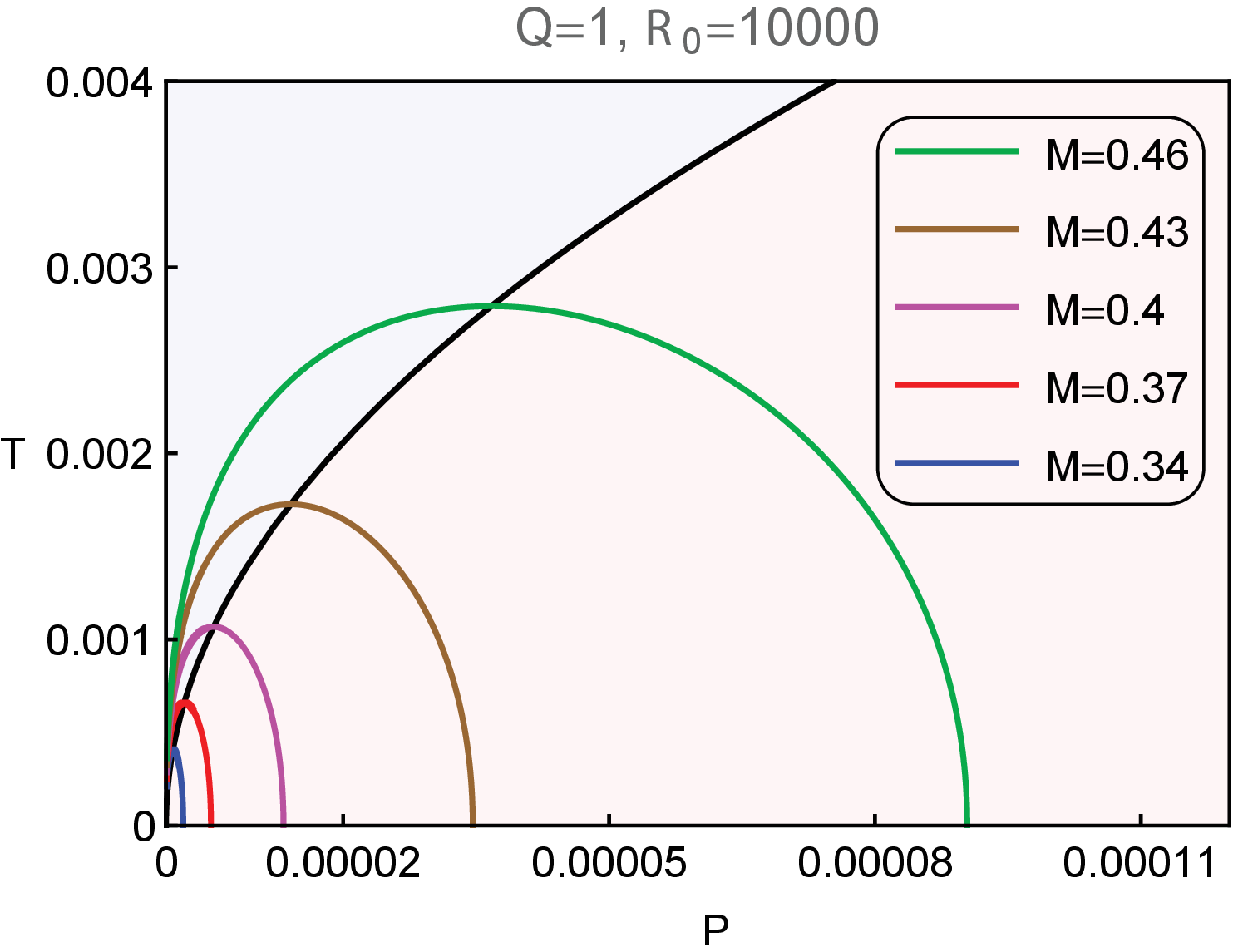}
	\hskip 0.5 cm
	\includegraphics[scale=0.37]{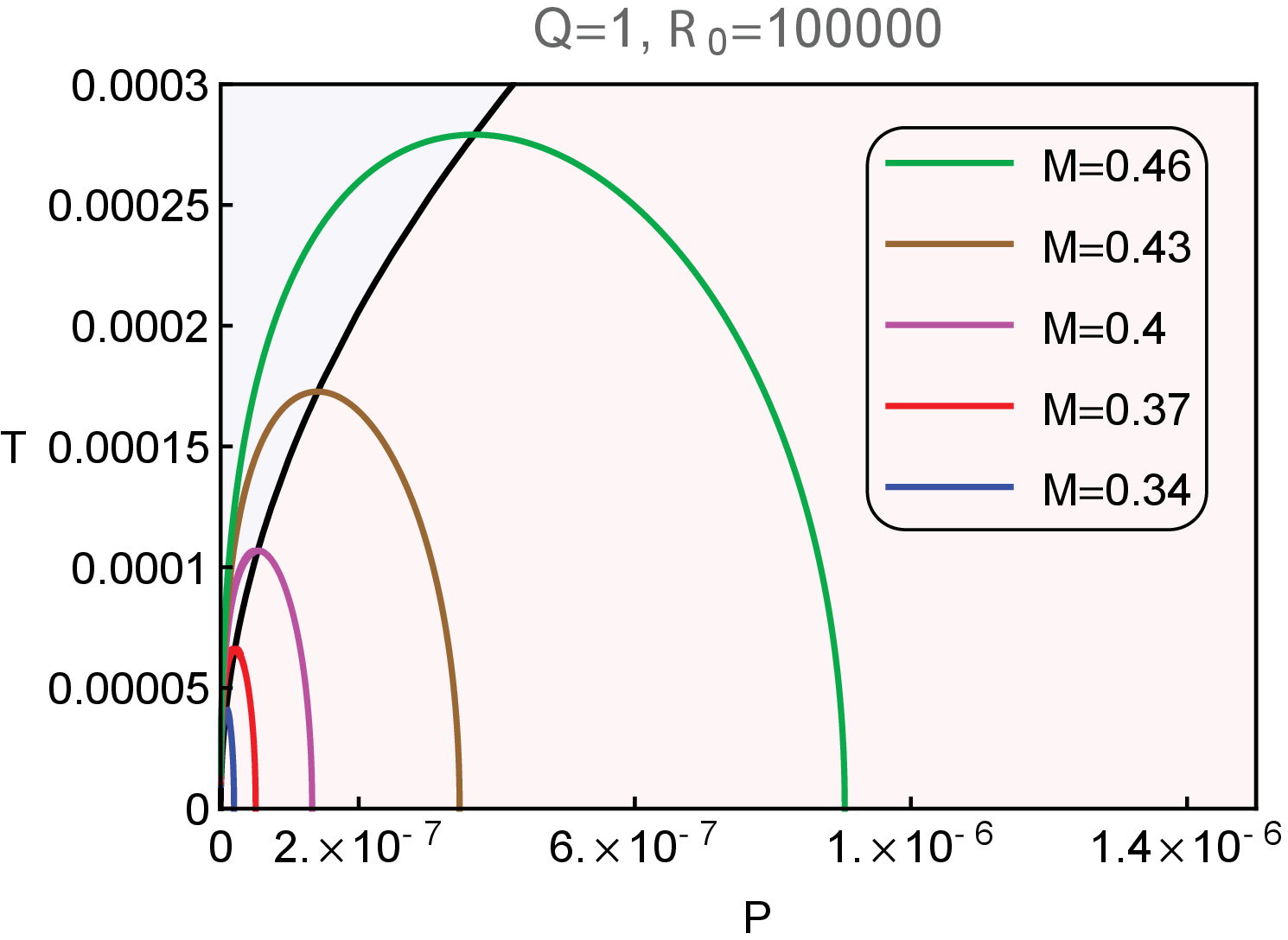}
	\caption{The inversion curves (black curves) and the isenthalps (colored curves) of charged BTZ black holes with $Q=1$ and different values of $R_0$.  For the blue regions, decreasing pressure at constant enthalpy leads to cooling ($\mu>0$) and for the red regions heating occurs as a result of reducing pressure while keeping the enthalpy constant ($\mu<0$).  }
	\label{char_isenthalps_2}
\end{figure}

Fig. \ref{char_inv_Q} shows the inversion curves for different values of $Q$ in the $T-P$ plane for the charged BTZ black holes. The behavior of inversion curves are qualitatively similar to that of rotating BTZ black holes, i.e., both are concave functions which admits there are no maximum inversion temperatures for both. Cooling occurs above the inversion curve so that if the initial temperature chosen is above the inversion curve, the final temperature is necessarily lower than the initial temperature, for fixed values of $P$ and $Q$ while holding the enthalpy constant.  \\
Solving Eq. (\ref{MCBH}) for $S$
\begin{equation}\label{a22}
S =  - \frac{i}{8}\sqrt {\frac{\pi }{2}} \frac{Q}{{\sqrt P }}\sqrt {{W_{ - 1}}\left( { - \frac{{32\pi R_0^2{{\rm{e}}^{ - \frac{{32M}}{{{Q^2}}}}}P}}{{{Q^2}}}} \right)} ,
\end{equation}
and substituting this entropy into the temperature expression in (\ref{rest}) results in
 \begin{equation}\label{TP2}
T =  - \frac{{{i}\sqrt P Q\left( {{W_{ - 1}}\left( { - \frac{{32\pi R_0^2{{\rm{e}}^{ - \frac{{32M}}{{{Q^2}}}}}P}}{{{Q^2}}}} \right) + 1} \right)}}{{\sqrt {2\pi } \sqrt {{W_{ - 1}}\left( { - \frac{{32\pi R_0^2{{\rm{e}}^{ - \frac{{32M}}{{{Q^2}}}}P}}}{{{Q^2}}}} \right)} }},
 \end{equation}
 where $W_{-1}$ denotes the Lambert W function and $i$ is the imaginary unit.\footnote{For the positive values of $P$ and $M$, the Lambert W function with the input argument as in Eqs. (\ref{a22})-(\ref{TP2}) is always negative and hence the entropy and the temperature will be real, as expected.} Using the above relation, the isenthalpic curves in the $T-P$ plane are plotted in Fig. \ref{char_isenthalps_1} for $R_{0}=100$ and $Q=0.8, 1$ and $1.2$. Fig. \ref{char_isenthalps_2} shows the effect of increasing $R_{0}$ on the isenthalps. Notice that the inversion curves are independent of $R_{0}$. As stated before, for the cooling region which is colored blue and lies above the inversion curve, the isenthalpic curves have positive slopes and for the red heating region, the slopes of the isenthalps are negative.    

At the end it is worth noting that while the equation of state of charged BTZ Black hole in scheme (I) is the same as that of scheme (II), i.e., the same as Eq. (\ref{eqosCBH}), and one might suggests that the Joule-Thomson expansion is possible for this system, nevertheless Eq. (\ref{PVTS}) contradicts this claim. Comparing (\ref{PVTS}) and (\ref{scbh}) with the equation of state of a non-interacting ideal gas affirms that there is no interaction in the charged BTZ black holes within scheme (I). Consequently, this scheme is self-incompatible, However, following any of the methods used for computing the inversion temperature yields $T_{i}=0$ for this scheme.

\section{Joule-Thomson expansion in charged rotating BTZ black holes}\label{JT in charged rotating BTZ}
\begin{figure}
	\includegraphics[scale=0.96]{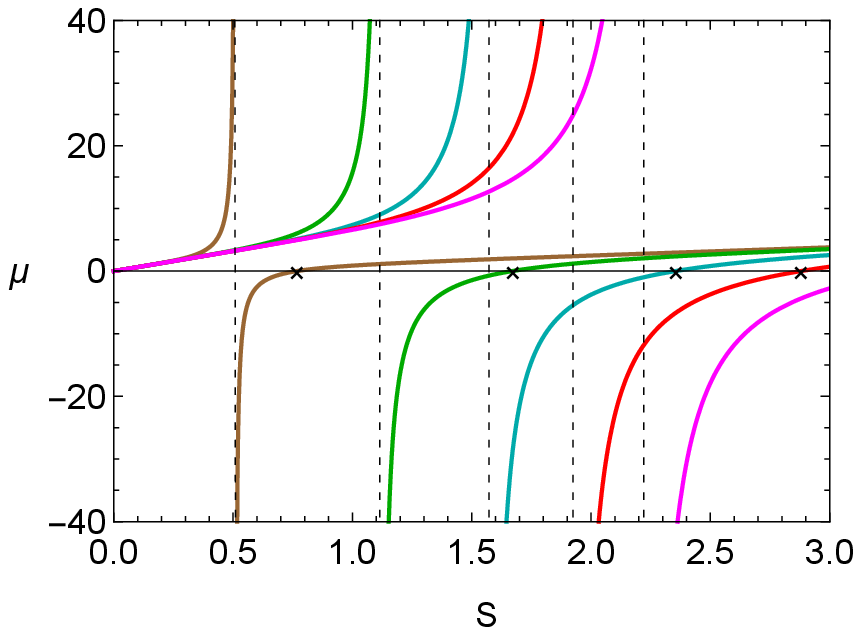}
	\hskip 1 cm
	\includegraphics[scale=0.9]{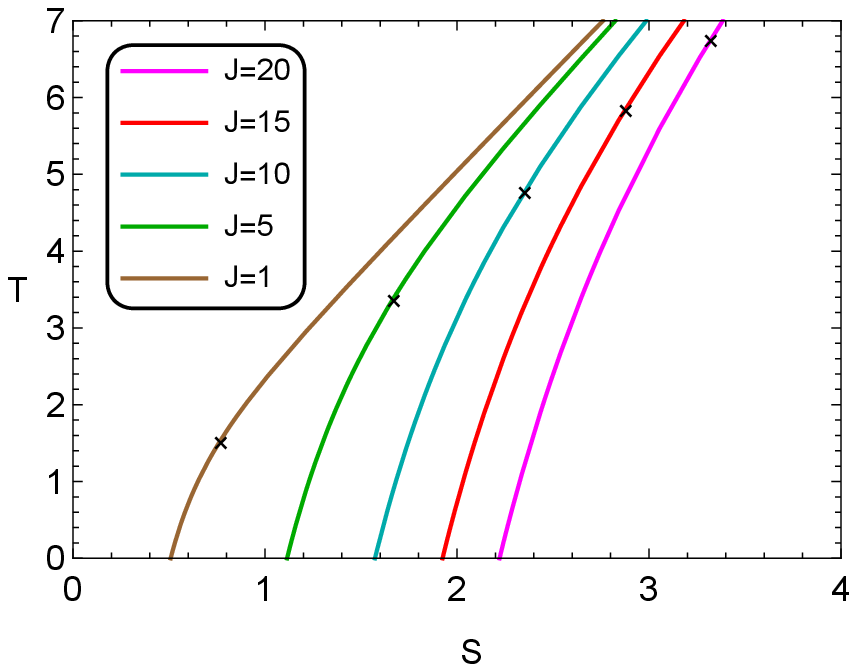}
	\caption{\textbf{$\mu$ vs $S$ (left panel) and $T$ vs $S$ (right panel) for charged rotating BTZ black holes}: Diagrams are plotted with fixed  $P=1$ and $Q=1$ and different values of $J$. The divergent points of the JT coefficient in the left panel (vertical dashed lines) are consistent with the zero point of the Hawking temperatures in the right panel. In the physical domain with positive temperature, $\mu$ increases monotonically and becomes zero only once for each value of $Q$ and $J$ (illustrated with cross marks). Therefore there are only minimum inversion temperatures for charged rotating BTZ black holes, i.e., the BTZ black holes always cool at large entropies $(\mu >0)$ and heat at small ones $(\mu <0)$ . The corresponding minimum inversion temperatures for different values of $Q$ are shown by cross marks in the right panel as well.}
	\label{char_rot_muTS}
\end{figure}

\begin{figure}
	\includegraphics[scale=0.35]{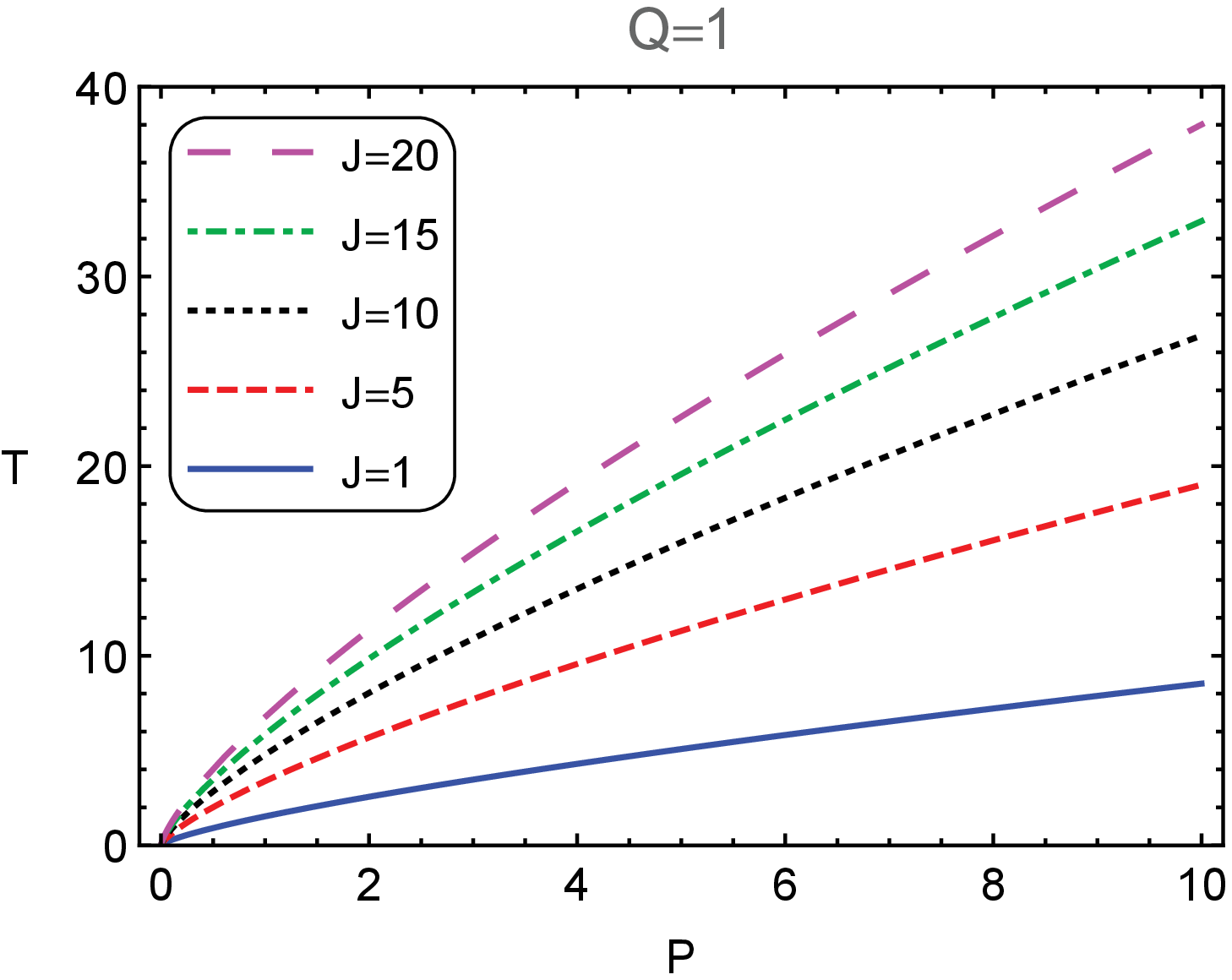}
	\hskip 0.5 cm
	\includegraphics[scale=0.35]{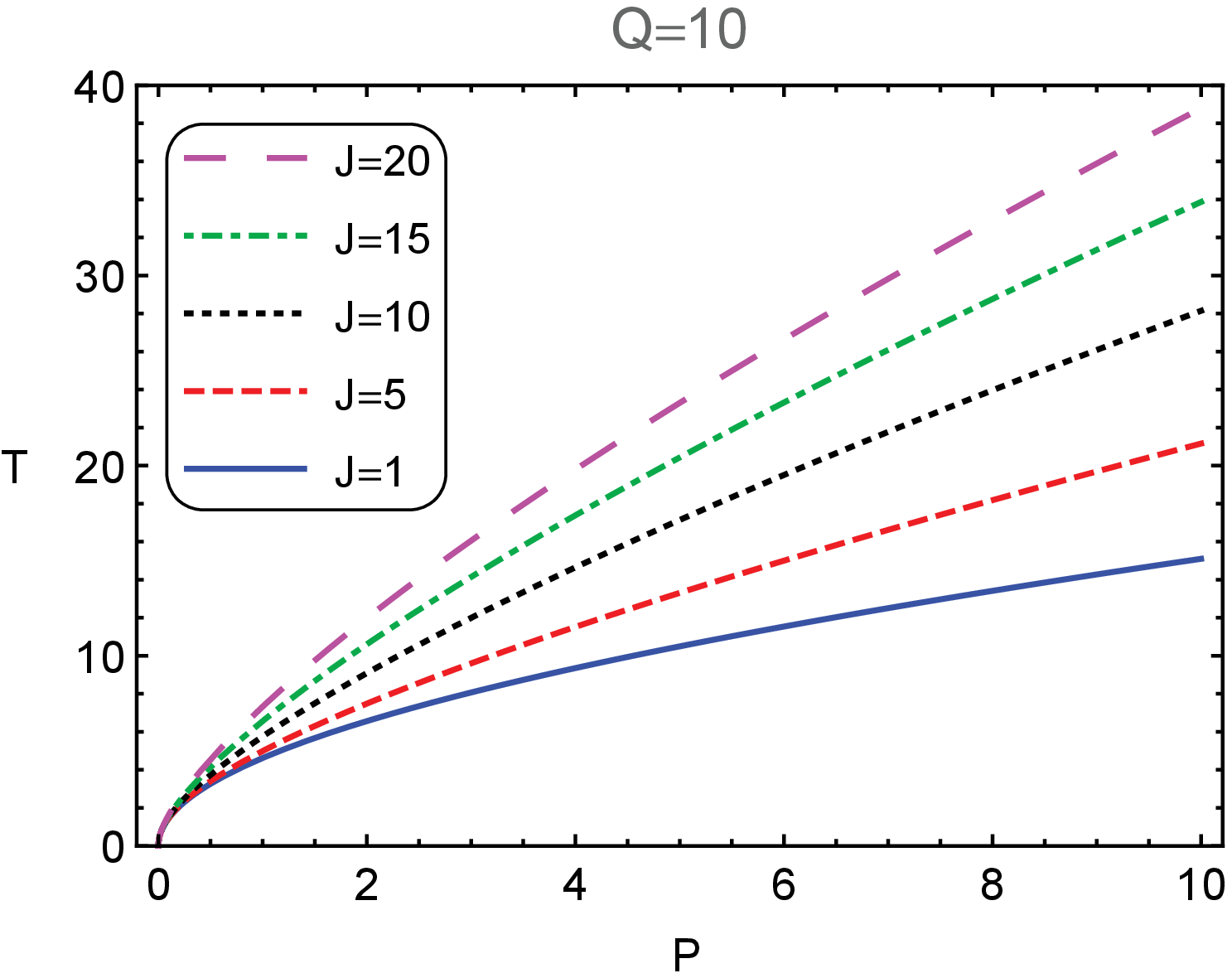}
	\hskip 0.5 cm
	\includegraphics[scale=0.35]{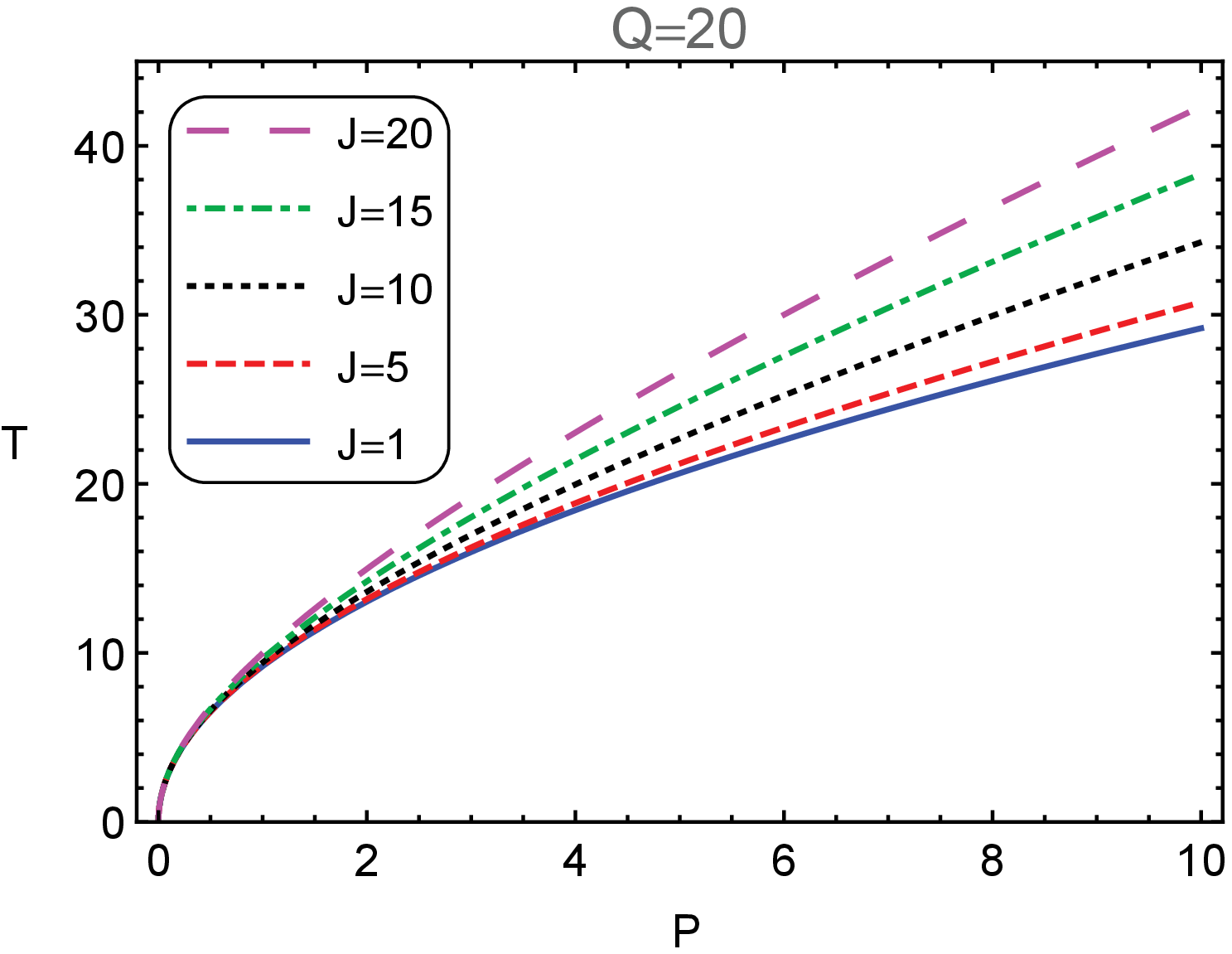}
	\caption{The inversion curves of the charged rotating BTZ black holes. From left to right, the inversion curves are plotted with fixed $Q=1$, $Q=10$, $Q=20$ and  five different values of $J$'s. Cooling and heating occurs above and below the inversion curve for each specified $J$ and $Q$. The inversion curves behave qualitatively in the same manner and they tend to converge for large $Q$'s.}
	\label{char_rot_inv_Q_J}
\end{figure}

\begin{figure}
	\includegraphics[scale=0.35]{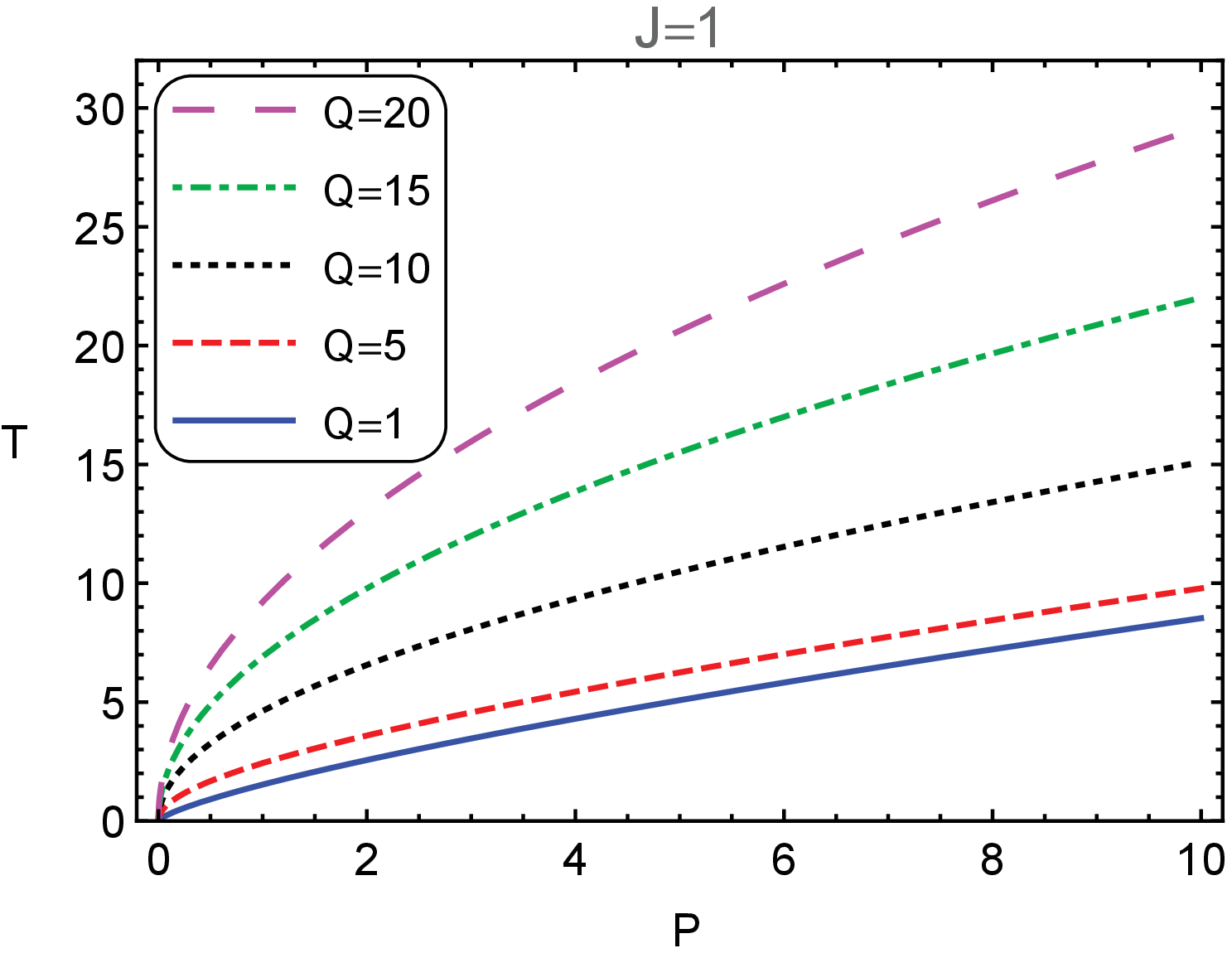}
	\hskip 0.5 cm
	\includegraphics[scale=0.35]{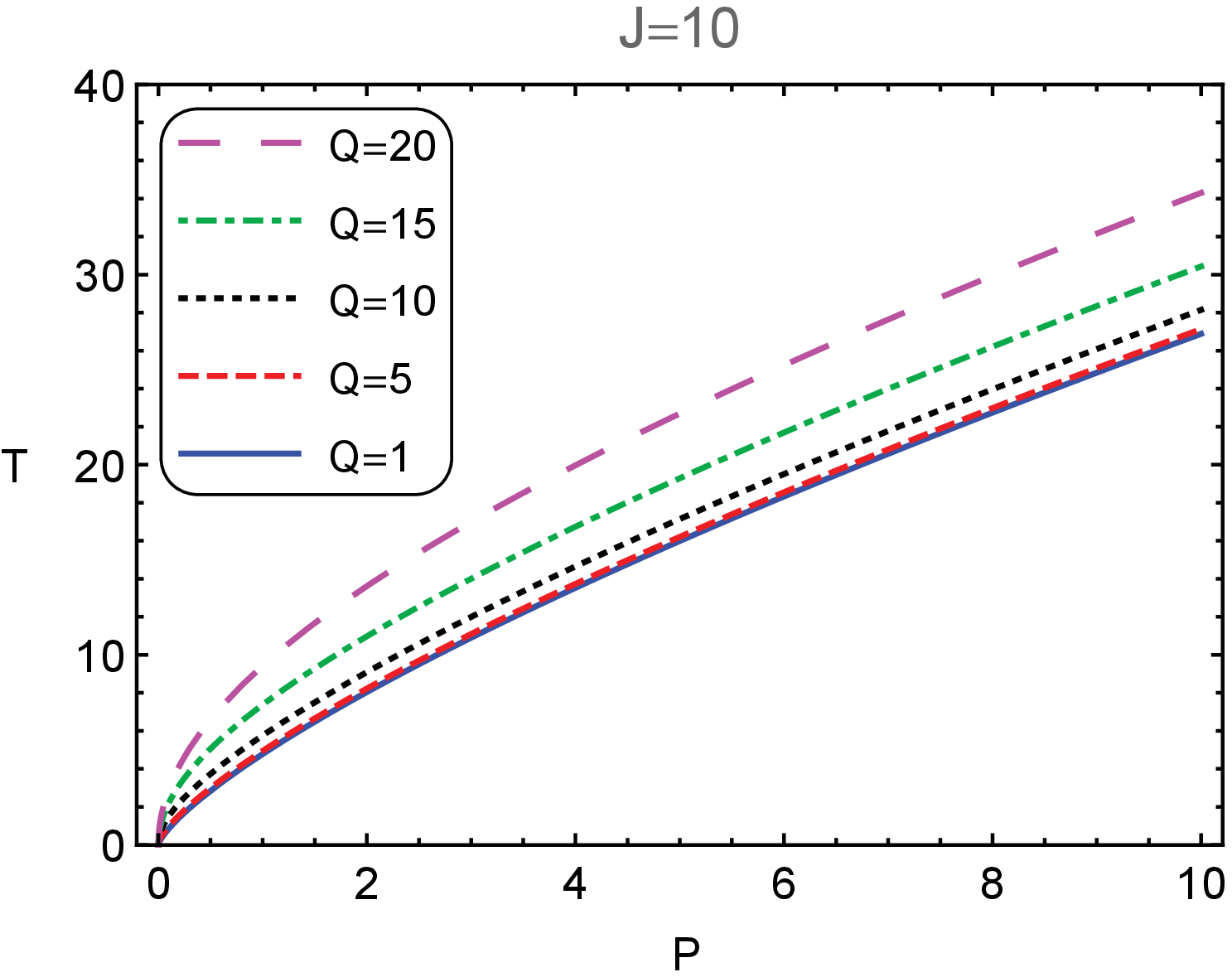}
	\hskip 0.5 cm
	\includegraphics[scale=0.35]{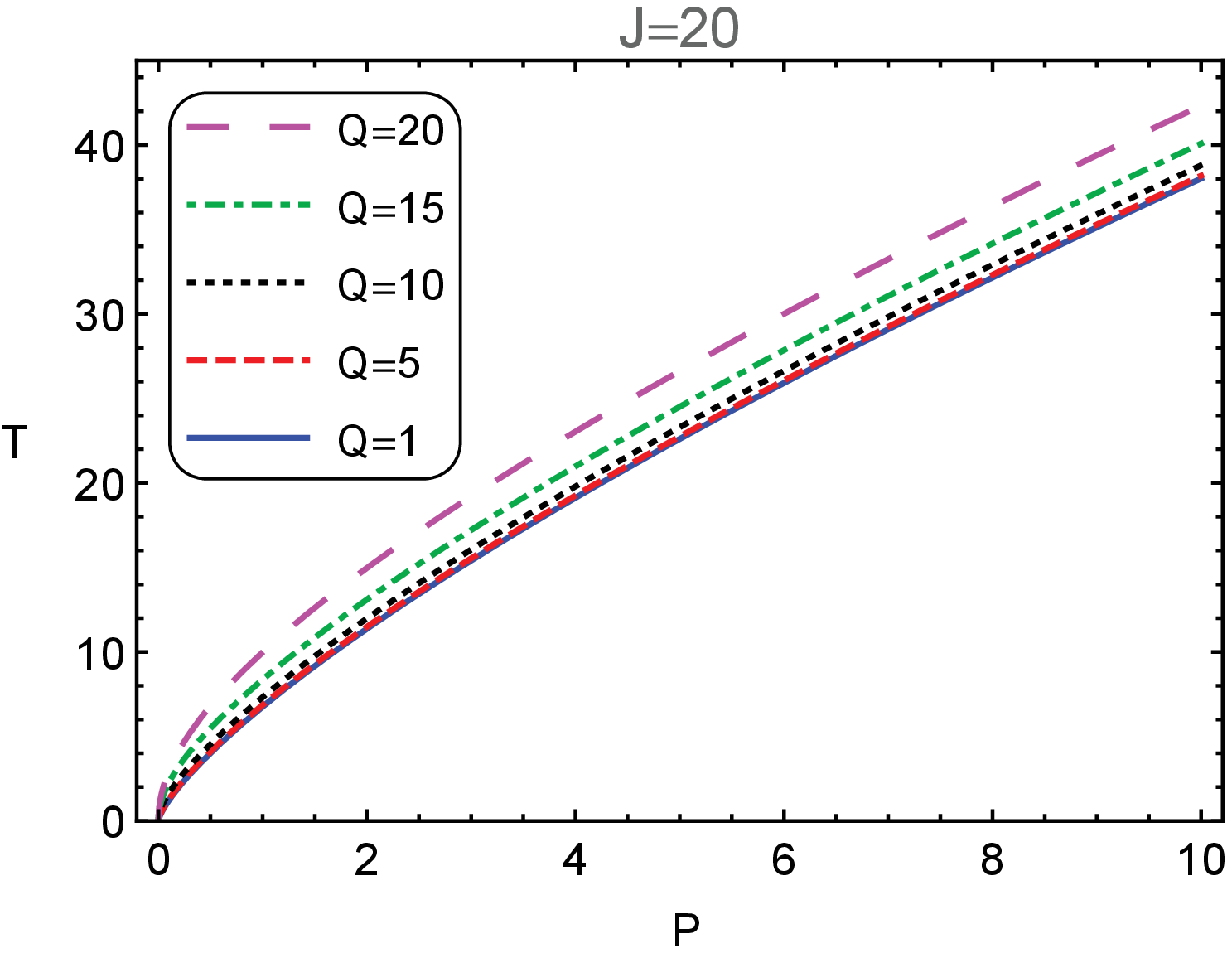}
	\caption{The inversion curves of the charged rotating BTZ black holes. From left to right, the inversion curves are plotted with fixed $J=1$, $J=10$, $J=20$ and  five different values of $Q$'s. Cooling and heating occurs above and below the inversion curve for each specified $J$ and $Q$. The inversion curves behave qualitatively in the same manner and they tend to converge for large $J$'s.}
	\label{char_rot_inv_J_Q}
\end{figure}

First, it should be clarified that for the charged rotating BTZ black holes, the same as charged BTZ black holes, the Joule-Thomson expansion will be examined within scheme II to avoid thermodynamical instabilities. Make use of Eq. (\ref{muu}), the Joule-Thomson coefficient for the charged rotating BTZ black hole reads
\begin{equation}
\mu  = 4S\left( {\frac{1}{\pi } + \frac{{4\left( {{J^2}{\pi ^2} + 2{Q^2}{S^2}} \right)}}{{{J^2}{\pi ^3} + 4\pi {Q^2}{S^2} - 512P{S^4}}}} \right).
\end{equation}
The left panel of Fig. \ref{char_rot_muTS} shows the behavior of the Joule-Thomson coefficient of the charged rotating BTZ black hole with fixed $P=1$, $Q=1$ and different values of $J$. As expected, similar to the cases of rotating and charged BTZ black holes, $\mu$ is an increasing function of $S$ and becomes zero only once which means that there are only minimum inversion temperatures (shown by cross marks in the right and left panel). Cooling and heating occurs at large and small entropies, respectively. \\   
Solving $\mu=0$, the corresponding inversion pressure and temperature are found
\begin{eqnarray}\label{invcr}
{P_i} &=& \frac{{5{J^2}{\pi ^3} + 12\pi {Q^2}{S^2}}}{{512{S^4}}},\nonumber\\
{T_i} &=& \frac{{256\pi {J^2}{P_i} + 6{Q^4} + 2{Q^2}\sqrt {640\pi {J^2}{P_i} + 9{Q^4}} }}{{\sqrt \pi  {P_i}{{\left( {\frac{{3{Q^2} + \sqrt {640\pi {J^2}{P_i} + 9{Q^4}} }}{{{P_i}}}} \right)}^{3/2}}}}.
\end{eqnarray}
Using the last line of Eq. (\ref{invcr}), the inversion curves for charged rotating BTZ black holes are plotted in Figs. \ref{char_rot_inv_Q_J} and \ref{char_rot_inv_J_Q}. For each specified value of $J$ and $Q$, cooling and heating occurs above and below the inversion curve. It is seen that the inversion curves behave qualitatively in the same manner and they tend to converge for large $Q$'s and $J$'s.

\begin{figure}
	\includegraphics[scale=0.37]{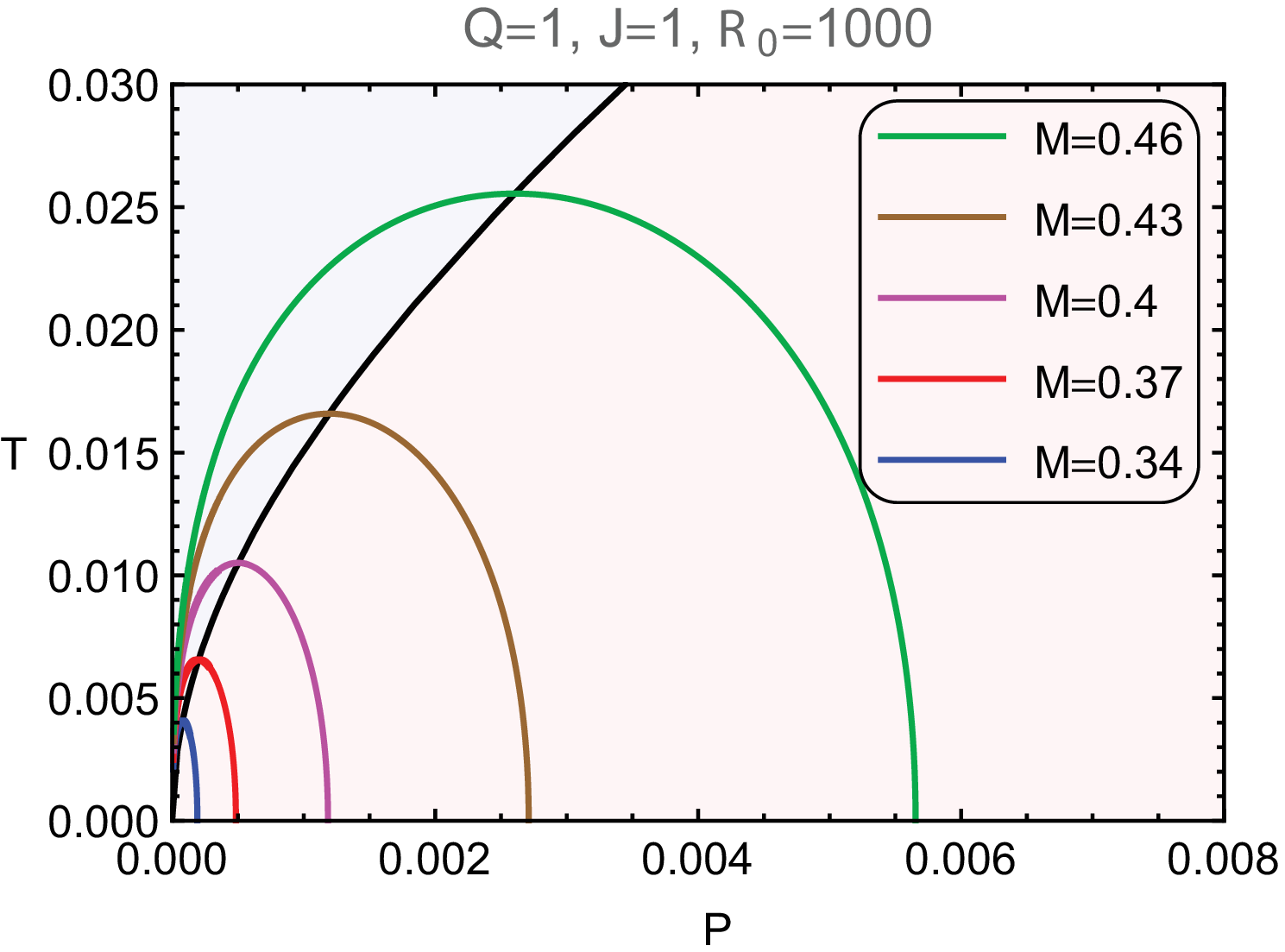}
	\hskip 0.5 cm
	\includegraphics[scale=0.37]{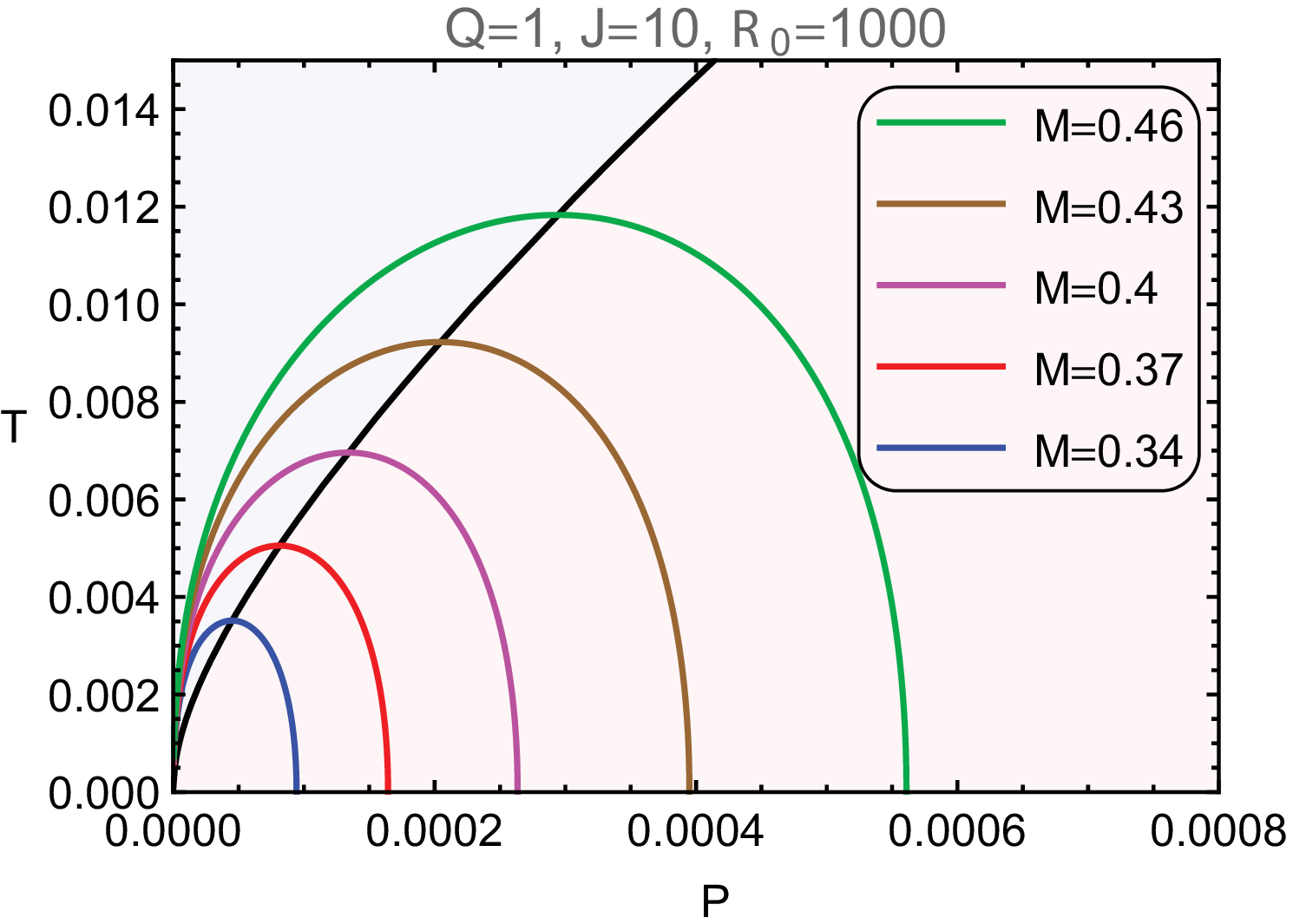}
	\hskip 0.5 cm
	\includegraphics[scale=0.35]{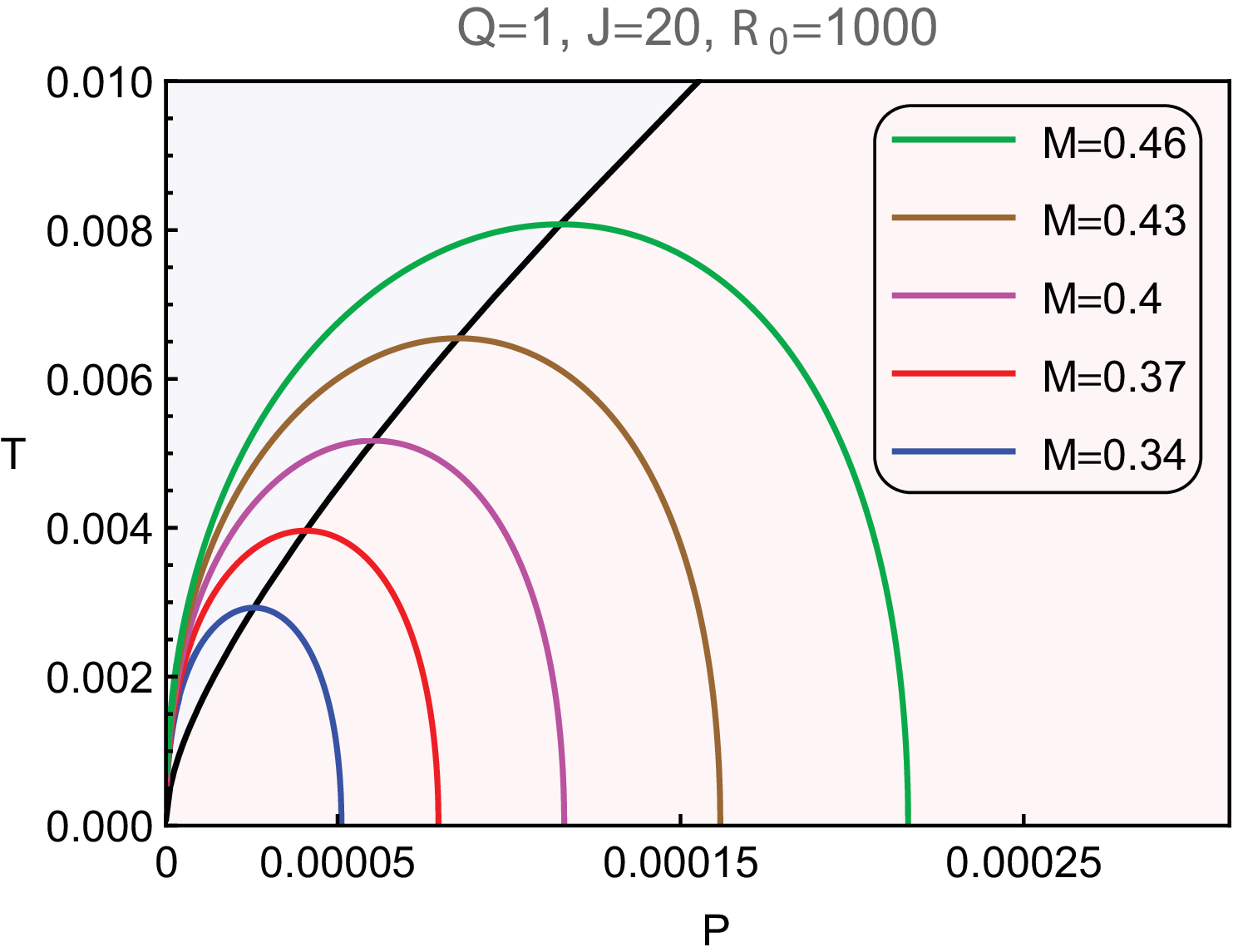}
	\vskip 0.5 cm
	\includegraphics[scale=0.35]{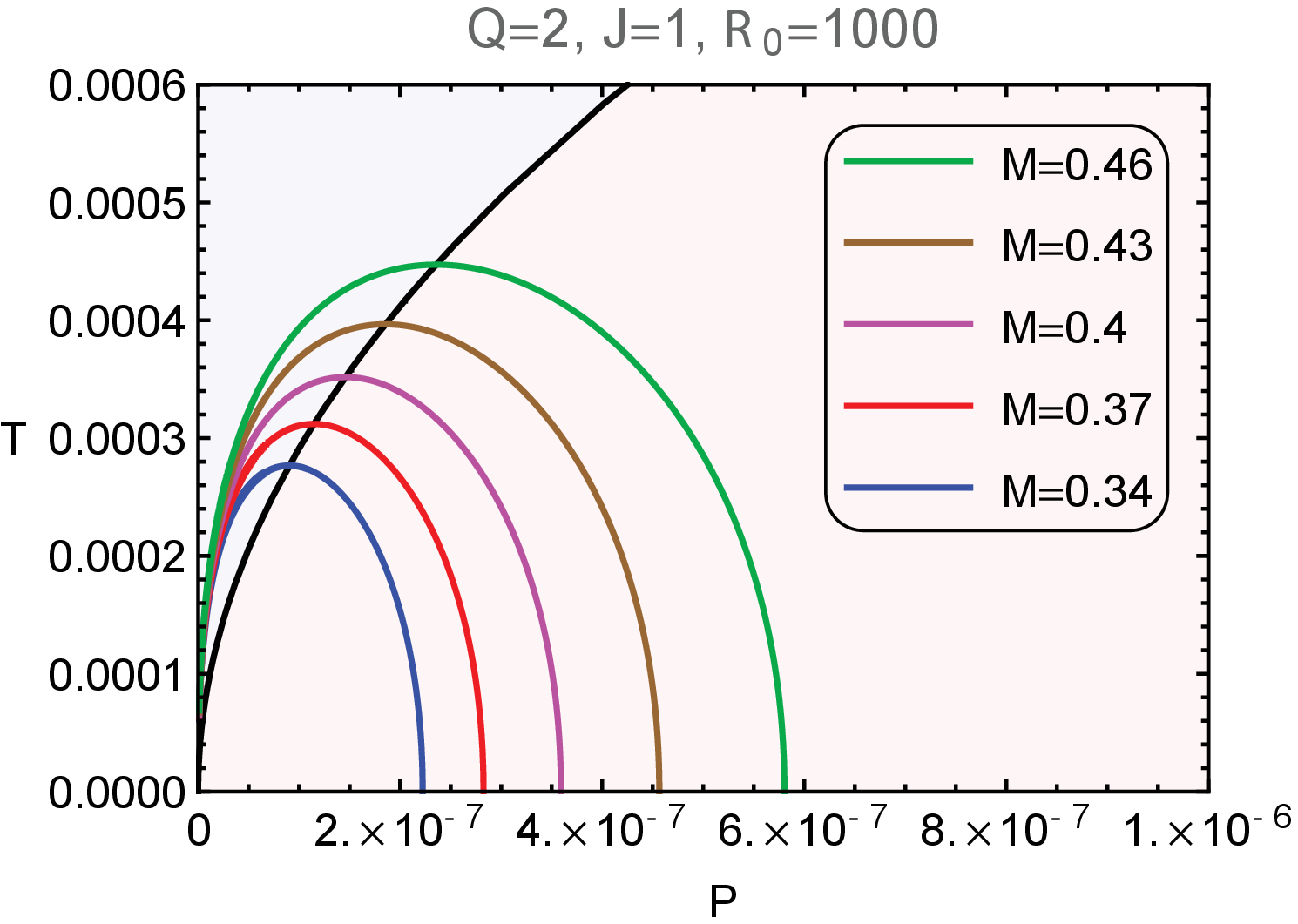}
	\hskip 0.5 cm
	\includegraphics[scale=0.35]{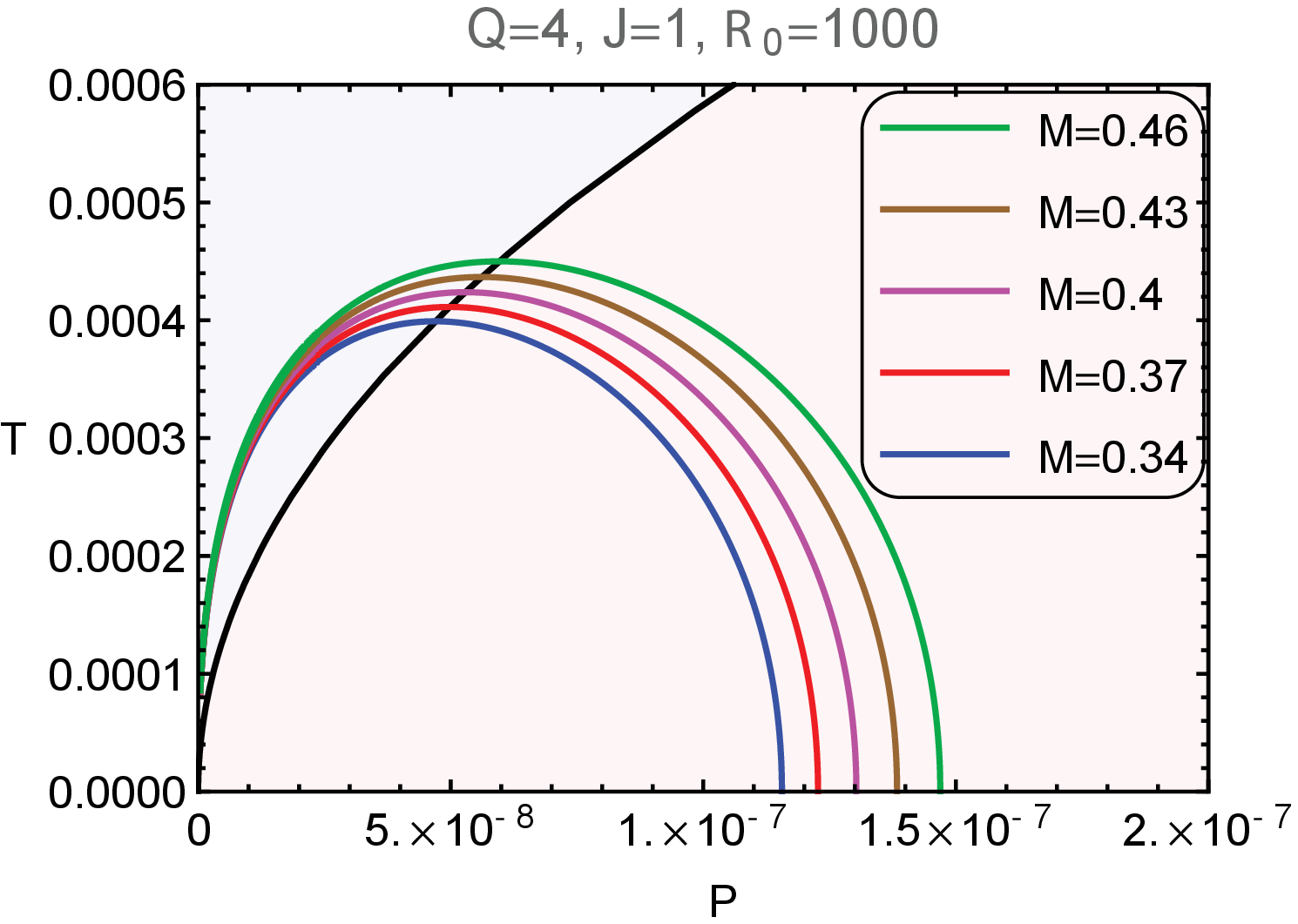}
	\caption{The inversion curves (black curves) and the isenthalps (colored curves) of charged rotating BTZ black holes with different values of $Q$ and $J$. For the blue regions, decreasing pressure at constant enthalpy leads to cooling ($\mu>0$) and for the red regions heating occurs as a result of reducing pressure while keeping the enthalpy constant ($\mu<0$).}
	\label{char_rot_isenthalps}
\end{figure}

Solving the first part of Eq. (\ref{MTcr}) analytically in order to obtain $r_{+}$ in terms of $M,P$ and $J$ is impossible and therefore the isenthalps for the charged rotating BTZ black holes are found by numerical methods. In Fig. (\ref{char_rot_isenthalps}) the isenthalpic curves besides the inversion curves are plotted with $R_0=1000$ for different values of $J$ and $Q$. The areas printed blue and red indicate cooling and heating regions, respectively.

\section{Closing remarks}\label{clore}
We performed the first study of holographic Joule-Thomson expansion in lower dimensions. For this, we considered three dimensional black hole spacetimes including both the charged and the rotating BTZ metrics and also BTZ metrics with both rotation and charge parameters. First, we studied the Joule-Thomson expansion for rotating BTZ black holes as the simplest case and it is verified that they are interacting systems (such a conjecture can be made from their equation of state, although for a precise scientific claim it is necessary to prove it.) We found the Joule-Thomson coefficient, the inversion temperature, the inversion curves in the $T-P$ plane and also the isenthalps. It was found that there are only minimum inversion temperatures and no maximum inversion temperatures as in the Van der Waals fluids.

\par
Next, we explored the Joule-Thomson expansion for the charged BTZ black holes in two possible schemes. In the traditional treatment, which is referred to as scheme I, the Joule-Thomson expansion does not occur due to its non-interacting nature which is obvious from the comparison of its Smarr relation with the equation of state of an ideal gas. In general, charged BTZ black holes in this scheme (I) are not appropriate for thermodynamical considerations since they always are thermally as well as mechanically unstable. Scheme II which is an alternative scheme, probably the correct one, for studying charged BTZ metrics possess a different set of thermodynamic relations and the Joule-Thomson expansion can occur for the charged BTZ black holes within this scheme. In this scheme, We investigated the Joule-Thomson coefficient, the inversion temperature and the inversion curves for this scheme. It was observed that the behavior of the inversion curves for charged BTZ black holes is qualitatively similar to those of rotating BTZ black holes. Furthermore, the analytic relation for isenthalpic curves was found. At last, we also examined the Joule-Thomson expansion for charged rotating BTZ black holes. The inversion curves were found to have the same qualitative manner as in the rotating and charged BTZ black holes.    

\par
As noted before, the Joule-Thomson expansion is absent in an ideal gas since there does not exist any interaction between its molecules. As an important conclusion, the existence of this process in 3D BTZ black holes indicates that it is possible to define the concept of black hole molecules with interactions.\footnote{Using the specific volume of AdS black holes, it is possible to define the concept of number density of black hole molecules as $n=1/v=(2 \ell_P^2 r_+)^{-1}$ to measure the microscopic degrees of freedom of the black hole \cite{Wei2015}. For more details about the concepts of black hole molecules and microscopic structure in AdS black holes see \cite{Wei2015}, in which the authors have discussed them completely from the thermodynamic viewpoint.} This is a very instructive subject since the number density of the BTZ black hole molecules has to be defined in two spatial dimensions, so we have a $2+1$ dimensional interacting system which cools and heats.

\par
In short, the most important findings are:
\begin{itemize}
	\item From a purely thermodynamic argument, the occurrence  of Joule-Thomson expansion in rotating, charged and charged rotating BTZ black holes explicitly indicates that these lower-dimensional black holes are interacting statistical systems.
	
	\item This outcome is not trivial for charged BTZ black holes, since it depends on the selected scheme. In the traditional scheme (I), the Joule-Thomson expansion does not occur which tells us that we are dealing with a noninteracting system. This is inconsistent with some other previous results \cite{Ghosh} in which the authors have explicitly shown that BTZ black holes in the presence of electric charge or angular momentum are associated with repulsive interactions among the microstructures (There exist other studies using thermodynamic geometry that indicate the thermodynamic interactions are present in charged or rotating BTZ black hole systems. For more details see Refs. \cite{TG2006,TG2009,TG2011}.) Within scheme II, the Joule-Thomson expansion arises which is a direct result of the fact that these systems are interacting in line with previous results \cite{TG2006,TG2009,TG2011,Ghosh}. 
	
\item Regarding the extended phase space of charged BTZ black holes, we have provided another piece of evidence indicating that scheme II is the favored scheme and the traditional scheme (I) is incompatible from a completely thermodynamic point of view. This is in support of the claim in \cite{Gregory2020} in which the authors stated that the traditional thermodynamics (in scheme I) of charged BTZ black holes is most likely not the correct one. This is an interesting result because, in scheme II, the variable $R_0$ has a natural interpretation. In fact, as discussed in Sec. \ref{CBH}, in order for renormalizing the total mass of the spacetime, it is necessary to enclose the system in a circle of radius $R_0$ \cite{Setare2008}. So, in scheme II, the area (a circle with radius $R_0$) enclosing the black hole spacetime changes as the energy content of system changes.
\end{itemize}
\par
There are some generalizations for both charged and rotating BTZ metrics in which this process can be further analyzed in lower dimensions. One can apply the Joule-Thomson expansion for generalized exotic BTZ black holes (as the solutions of the exotic 3D Einstein gravity \cite{ExoticBTZ2013Townsend}). But these solutions suffers from a fundamental thermodynamic instability as shown in Refs. \cite{Johnson2020MPLA,EBTZinstability2019}. Furthermore, we predict that a nonlinear generalization of charged BTZ black holes, e.g., within the context of Einstein-Euler-Heisenberg theory at weak-field limit (${{}\cal L}_\text{EEH}=R-2\Lambda-{\cal F}+\gamma{\cal F}^2$), also could exhibit this expansion process but one has to take care of the particular scheme. In the charged BTZ scheme I, the black hole solutions can be thermodynamically stable in a small region of phase space but the corresponding behavior of the electric potential at asymptotic region is still ill-defined. So, in this case, the BTZ scheme II is also the preferred choice. Having this, the effect of quadratic nonlinear electromagnetic (radiative) corrections in charged BTZ black holes can be analyzed for the process of Joule-Thomson expansion.

\begin{acknowledgements}
Wewould like to thank the anonymous referee who provided useful and detailed comments on a previous version of the manuscript. S.Z. would like to express her sincere gratitude to A. Dehghani for useful discussions. S.Z. also appreciates the
support of Sistan and Baluchestan University research council.
\end{acknowledgements}



\begin{thebibliography} {0}
	%
	
	\bibitem{beck1}
	J.D. Bekenstein, \textit{Black holes and the second law}, Lett. Nuovo Cimento \textbf{4} (1972) 737.
	
	\bibitem{beck2}
	J.D. Bekenstein, \textit{Black Holes and Entropy}, Phys. Rev. D \textbf{7} (1973) 2333.
	
	\bibitem{hawk1}
    S.W. Hawking,  \textit{Black hole explosions?}, Nature \textbf{248} (1974) 30.
    
    \bibitem{hawk2}
    S.W. Hawking, \textit{Particle creation by black holes}, Commun. Math. Phys. 43 (1975) 199.
    
	\bibitem{hawk3}
	S. W. Hawking, \textit{Black holes and thermodynamics}, Phys. Rev. D \textbf{13} (1976) 191.
	
	
	\bibitem{HawkingPage1983}
	S. Hawking and D.N. Page, \textit{Thermodynamics of black holes in Anti-de Sitter space}, Commun. Math. Phys. 87 (1983) 577.
	 
	 \bibitem{Myers1988}
	 R.C. Myers and J.Z. Simon, Black-hole thermodynamics in Lovelock gravity, Phys. Rev. D 38 (1988) 2434
	 
	 \bibitem{Chamblin1}
	 A. Chamblin, R. Emparan, C.V. Johnson, and R.C. Myers, \textit{Charged AdS black holes and catastrophic holography}, Phys.
	 Rev. D \textbf{60} (1999) 064018.
	 
	 
	 \bibitem{Chamblin2}
	 A. Chamblin, R. Emparan, C.V. Johnson, and R.C. Myers, Holography, \textit{thermodynamics and 
uctuations of charged AdS
	 black holes}, Phys. Rev. D \textbf{60} (1999) 104026.
	 
	
	 \bibitem{Fernando}
	 S. Fernando, \textit{Thermodynamics of Born-Infeld-anti-de Sitter black holes in the grand canonical ensemble}, Phys. Rev. D \textbf{74}
	 (2006) 104032.
	 

    \bibitem{Henneaux1984}
    M. Henneaux and C. Teitelboim, \textit{The cosmological constant as a canonical variable}, Phys. Lett. B \textbf{143} (1984) 415.
    
   \bibitem{Teitelboim}
    C. Teitelboim, \textit{The cosmological constant as a thermodynamic black hole parameter}, Phys. Lett. B \textbf{158} (1985) 293.
    
    \bibitem{Henneaux1989}
    M. Henneaux and C. Teitelboim, \textit{The cosmological constant and general covariance}, Phys. Lett. B \textbf{222} (1989) 195.
    
    \bibitem{cald}
      M. M. Caldarelli, G. Cognola and D. Klemm, \textit{Thermodynamics of Kerr-Newman-AdS black holes and conformal field theories}, Classical and Quantum Gravity \textbf{17} (2000) 399.
      
      
      \bibitem{shuang}
      W. Shuang, W. Shuang-Qing, X. Fei and D. Lin, \textit{The First Law of Thermodynamics of the $(2+1)$-Dimensional Banados-Teitelboim-Zanelli Black Holes and Kerr-de Sitter Spacetimes}, Chinese Physics Letters \textbf{23} (2006) 1096.
	
	\bibitem{Sekiwa}
	Y. Sekiwa, \textit{Thermodynamics of de Sitter black holes: thermal cosmological constant}, Physical Review D \textbf{73} (2006) 084009.
	
	
	\bibitem{Kastor2009} D. Kastor, S. Ray, and J. Traschen, \textit{Enthalpy and the mechanics of AdS black holes}, Class. Quant. Grav. \textbf{26} (2009) 195011.
	
	\bibitem{Dolan1} B.P. Dolan, \textit{The cosmological constant and the black hole equation of state}, Class. Quant. Grav. \textbf{28} (2011) 125020.
	
	\bibitem{Dolan2}B.P. Dolan, \textit{Pressure and volume in the first law of black hole thermodynamics}, Class. Quant. Grav. \textbf{28} (2011) 235017.
	
	\bibitem{Kubiznak2012}
	D. Kubiznak and R.B. Mann, \textit{P-V criticality of charged AdS black holes}, JHEP \textbf{07 }(2012) 033.
	
	\bibitem{Bazanski1990}
	S.L. Bazanski and P. Zyla, \textit{A Gauss type law for gravity with a cosmological constant}, Gen. Rel. Grav. \textbf{22} (1990) 379.
	
	
	\bibitem{Kubiznak2017}
	D. Kubiznak, R.B. Mann and M. Teo, \textit{Black hole chemistry: thermodynamics with Lambda}, Class. Quant. Grav. \textbf{34} (2017) 063001.
	
	\bibitem{Gunasekaran}
	S. Gunasekaran, R.B. Mann, and D. Kubiznak, \textit{Extended phase space thermodynamics for charged and rotating black holes and Born-Infeld vacuum polarization}, JHEP \textbf{11} (2012) 110.
	
	\bibitem{Astefanesei2019}
	D. Astefanesei, R.B. Mann and R. Rojas, \textit{Hairy black hole chemistry}, JHEP \textbf{11} (2019) 043.
	
	
	\bibitem{Johnson2014} 
	C.V. Johnson, \textit{Holographic heat engines}, Class. Quant. Grav. \textbf{31} (2014) 205002
	
	\bibitem{Johnson2016a} 
	C.V. Johnson, \textit{Born-Infeld AdS black holes as heat
	engines}, Class. Quant. Grav. \textbf{33} (2016) 135001
	
	\bibitem{Johnson2016b}
	 C.V. Johnson, \textit{Gauss-Bonnet black holes and holographic heat engines beyond large \textit{N}}, Class. Quant. Grav. \textbf{33} (2016) 215009
	
	
	\bibitem{HE2019_3Dim} 
	L. Balarta and S. Fernando, \textit{Non-linear black holes in 2+1 dimensions as heat engines}, Phys. Lett. B \textbf{795} (2019) 638
	
	\bibitem{HE2017Hennigar} 
	R.A. Hennigar, F. McCarthy, A. Ballon, and R.B. Mann, \textit{Holographic heat engines: general considerations and rotating black holes}, Class. Quant. Grav. \textbf{34} (2017) 175005
	
	\bibitem{HE2018}
	 Jie-Xiong Mo and Gu-Qiang Li, Holographic heat engine within the framework of massive gravity, JHEP \textbf{05} (2018) 122
	
	\bibitem{HE2021SZ}
	S. Zarepour, \textit{Holographic heat engines coupled with logarithmic $U(1)$ gauge theory}, To be published in Int. J. Mod. Phys. D, https://doi.org/10.1142/S0218271821501091
	
	
	
	
	\bibitem{JT2017a} 
	\"{O}. \"{O}kc\"{u} and E. Aydıner, \textit{Joule-Thomson expansion of the charged AdS black holes}, Eur. Phys. J. C \textbf{77} (2017) 24.
	
	\bibitem{JT2018a}
	 \"{O}. \"{O}kc\"{u} and E. Aydıner, \textit{Joule-Thomson expansion of Kerr-AdS black holes}, Eur. Phys. J. C \textbf{78} (2018) 123.
	
	\bibitem{JT2018c} 
	A. Haldar and R. Biswas, \textit{Joule-Thomson expansion of five-dimensional Einstein-Maxwell-Gauss- Bonnet-AdS black holes}, EPL \textbf{123} (2018) 40005.
	
	\bibitem{JT2018h}
	C.L.A. Rizwan, A.N. Kumara, D. Vaid, and K.M. Ajith, \textit{Joule-Thomson expansion in AdS black hole with a global monopole}, Int. J. Mod. Phys. A \textbf{33} (2018) 1850210.
	
	
	\bibitem{JT2018i}
	H. Ghaffarnejad, E. Yaraie, and M. Farsam, \textit{Quintessence Reissner-Nordstr\"om anti de Sitter black holes and Joule Thomson effect}, Int. J. Theor. Phys. \textbf{57} (2018) 1671.
		
	\bibitem{JT2018d}
	M. Chabab, H. El Moumni, S. Iraoui, Masmar, and S. Zhizeh, \textit{Joule-Thomson Expansion of RN-AdS Black Holes in $f(R)$ gravity}, LHEP \textbf{02} (2018) 05.	
		
	\bibitem{JT2018b} 
	J-X. Mo, G-Q. Li, S-Q. Lan, and X-B. Xu, \textit{Joule-Thomson expansion of $ d $-dimensional charged AdS black holes}, Phys. Rev. D \textbf{98} (2018) 124032. 
	
	\bibitem{JT2018g} 
	S-Q. Lan, \textit{Joule-Thomson expansion of charged Gauss-Bonnet black holes in AdS space}, Phys. Rev. D \textbf{98} (2018) 084014.
	
	\bibitem{JT2018e} 
	Z-W. Zhao, Y-H. Xiu, and N. Li, \textit{Throttling process of the Kerr-Newman-anti-de Sitter black holes in the extended phase space}, Phys. Rev. D \textbf{98} (2018) 124003.
	
    \bibitem{JT2020g}
	C. Li, P. He, P. Li, and J-B. Deng, \textit{Joule-Thomson expansion of the Bardeen-AdS black holes}, Gen. Rel.  Grav. \textbf{52} (2020) 1. 
	
	\bibitem{JT2019a} 
	S-Q. Lan, \textit{Joule-Thomson expansion of neutral AdS black holes in massive gravity}, Nuc. Phys. B \textbf{948} (2019) 114787.
	
	\bibitem{JT2018f}
	X-M. Kuang, B. Liu, and Ali \"{O}vg\"{u}n, \textit{Nonlinear electrodynamics AdS black hole and related phenomena in the extended thermodynamics}, Eur. Phys. J. C \textbf{78} (2018) 840.
	
	
	\bibitem{JT2019b}
	 A. Cisterna, S-Q. Hub, X-M. Kuang, \textit{Joule-Thomson expansion in AdS black holes with momentum relaxation}, Phys. Lett. B \textbf{797} (2019) 134883.
	
	\bibitem{JT2019c}
	 D. Mahdavian Yekta, A. Hadikhani, \"{O}. \"{O}kc\"{u}, \textit{Joule-Thomson expansion of charged AdS black holes in Rainbow gravity}, Phys. Lett. B \textbf{795} (2019) 521.
	
	\bibitem{JT2020a} 
	S. Guo,Y. Han, and G-P. Li, \textit{Joule-Thomson expansion of a specific black hole in $ f(R) $ gravity coupled with Yang-Mills field}, Class. Quant. Grav. \textbf{37} (2020) 085016.
	
	\bibitem{JT2020b} 
	J. Sadeghi and R. Toorandaz, \textit{Joule-Thomson expansion of hyperscaling violating black holes with spherical and hyperbolic horizons}, Nuc. Phys. B \textbf{951} (2020) 114902.
	
	\bibitem{JT2020c}
	 C.H. Nam, \textit{Effect of massive gravity on Joule-Thomson expansion of the charged AdS black hole}, Eur. Phys. J. Plus \textbf{135} (2020) 259.
	
	\bibitem{JT2020d}
	 Y. Meng, Q-Q. Jiang, and J. Pu, \textit{The P-V criticality and Joule-Thomson expansion of charged AdS black hole in the Rastall gravity}, Chinese Phys. C \textbf{44} (2020) 065105.
	
	\bibitem{JT2020e}
	A. Jawad, M. Yasir, and S. Rani, \textit{Joule-Thomson expansion and quasinormal modes of regular non-minimal magnetic black hole}, Mod. Phys. Lett. A \textbf{35} (2020) 2050298.
	
	\bibitem{JT2020f}
	J-X. Mo and G-Q. Li, \textit{Effects of Lovelock gravity on the Joule–Thomson expansion}, Class. Quant. Grav. \textbf{37} (2020) 045009.
	
	\bibitem{JT2020h}
	 U. Debnath, \textit{Thermodynamics of FRW Universe: Heat engine}, Phys. Lett. B \textbf{810} (2020) 135807.
	
	\bibitem{JT2020i}
	 M. Rostami, J. Sadeghi, S. Miraboutalebi, A.A. Masoudi, and B. Pourhassan, \textit{Charged accelerating AdS black hole of $f (R)$ gravity and the Joule-Thomson expansion}, Int. J. Geom. Methods Mod. Phys. \textbf{17} (2020) 2050136.
	
	\bibitem{JT2020j}
	S. Guo, J. Pu, Q. Jiang and X. Zu, \textit{Joule-Thomson expansion of the regular(Bardeen)-AdS black hole}, Chinese Phys. C \textbf{44} (2020) 035102.
	
	\bibitem{JT2020k}
	S. Guo, Y. Han and GP. Li, \textit{Thermodynamic of the charged AdS black holes in Rastall gravity: $P-V$ critical and Joule–Thomson expansion}, Modern Physics Letters A, \textbf{35(14}) (2020) 2050113. 
	
	
	\bibitem{JT2021}
	 S. Bi, M. Du, J. Tao, and F. Yao, \textit{Joule-Thomson expansion of Born-Infeld AdS black holes}, Chinese Phys. C \textbf{45} (2021) 025109.
	
	
	
	
	\bibitem{arxiv1}
	S. Guo, Y. Huang, and E. Liang, \textit{Comparison of thermodynamic behavior of two regular-AdS black holes}, arXiv: 2009.03519 (2020).
	
	\bibitem{arxiv2}
	 M. Farsam, E. Yaraie, H. Ghaffarnejad, and E. Ghasami, \textit{Cooling-heating phase transition for 4D AdS Bardeen Gauss-Bonnet Black Hole}, arXiv: 2010.05697 (2020).
	
	\bibitem{arxiv3}
	 S. Guo, Y. Han, and G. Li, \textit{Joule-Thomson expansion of a specific black hole in different dimensions}, arXiv: 1912.09590 (2019).
	
	\bibitem{arxiv4}
	Z-W. Feng, X. Zhou, and S-Q. Zhou, \textit{Joule-Thomson expansion of higher dimensional nonlinearly charged AdS black hole in Einstein-PMI gravity}, Commun. Theor. Phys. \textbf{73} (2021) 065401
	
	\bibitem{arxiv5}
	Y. Huang and S. Guo, \textit{Thermodynamic of the charged accelerating AdS black hole: PV critical and Joule-Thomson expansion}, arXiv: 2009.09401 (2020).
	
	\bibitem{arxiv6} 
	K. Hegde, A. N. Kumara, C.L. Ahmed Rizwan, Ajith K.M., Md S. Ali, \textit{Thermodynamics, Phase Transition and Joule Thomson Expansion of novel 4-D Gauss Bonnet AdS Black Hole}, 	arXiv:2003.08778 (2020).
	
	\bibitem{arxiv7}
	 N. Chen, \textit{Throttling Process of Rotating Bardeen AdS Black Holes},  arXiv: 2003.00247 (2020).
	
	\bibitem{arxiv2021} 
	Y. Cao, H. Feng, W. Hong, and J. Tao, \textit{Joule-Thomson Expansion of RN-AdS Black Hole Immersed in Perfect Fluid Dark Matter}, Commun. Theor. Phys. \textbf{73} (2021) 095403.
		
	
	\bibitem{BTZ1992}
	  M.~Ba{\~{n}}ados, C.~Teitelboim and J.~Zanelli, \textit{Black hole in three-dimensional spacetime}, Phys. Rev. Lett. {\bf 69} (1992) 1849.
	
	\bibitem{BHTZ1993}
	  M.~Ba{\~{n}}ados, M.~Henneaux, C.~Teitelboim and J.~Zanelli, \textit{Geometry of the 2+1 black hole}, Phys. Rev. D \textbf{48} (1993) 1506.
	
	\bibitem{MTZ2000}
	C. Mart\'{\i }nez, C. Teitelboim, and J. Zanelli, \textit{Charged rotating black hole in three spacetime dimensions}, Phys. Rev. D \textbf{61} (2000) 104013.
	
	\bibitem{Ghosh}
	A. Ghosh and C. Bhamidipati, \textit{Thermodynamic geometry and interacting microstructures of BTZ black holes}, Phys. Rev. D \textbf{101} (2020) 106007.
	
	\bibitem{Liang}
	J. Liang, B. Mu and P. Wang, \textit{Joule-Thomson expansion of Lower-dimensional black hole}, 	arXiv:2104.08841 [gr-qc] (2021).
	
	
	\bibitem{Frassino2015} 
	A.M. Frassino, R.B. Mann, and J.R. Mureika, \textit{Lower-dimensional black hole chemistry}, Phys. Rev. D \textbf{92} (2015) 124069.
	
	\bibitem{Setare2008}
	M. Cadoni, M. Melis, and M.R. Setare. \textit{Microscopic entropy of the charged BTZ black hole}, Class. Quant. Grav. \textbf{25} (2008) 195022.
	
	\bibitem{Johnson2020MPLA}
	C.V. Johnson, \textit{Instability of super-entropic black holes in extended thermodynamics}, Mod. Phys. Lett. A \textbf{35} (2020) 2050098.
	
	\bibitem{TG2006}T. Sarkar, G. Sengupta, and B.N. Tiwari, \textit{On the thermodynamic geometry of BTZ black holes}, JHEP \textbf{11} (2006) 015.
	
	\bibitem{TG2009}H. Quevedo and A. Sanchez, \textit{Geometric description of BTZ black hole thermodynamics}, Phys. Rev. D \textbf{79} (2009) 024012.
	
	\bibitem{TG2011} M. Akbar, H. Quevedo, K. Saifullah, A. Sanchez, and Safia Taj, \textit{Thermodynamic geometry of charged rotating BTZ black holes}, Phys. Rev. D \textbf{8} (2011) 084031.
		
	
	\bibitem{Gregory2020}
	M. Appels, L. Cuspinera, R. Gregory, P. Krtous, and D. Kubiznak, \textit{Are “Superentropic” black holes superentropic?}, JHEP \textbf{02} (2020) 195.
	
	\bibitem{Wei2015}S.W. Wei and Y.X. Liu, \textit{Insight into the microscopic structure of an AdS black hole from a thermodynamical phase transition}, Phys. rev. lett. \textbf{115} (2015) 111302.
	
	\bibitem{ExoticBTZ2013Townsend}
	P. K. Townsend and B. Zhang, \textit{Thermodynamics of Exotic Ba{\~{n}}ados-Teitelboim-Zanelli Black Holes}, Phys. Rev. Lett. \textbf{110} (2013) 241302.
	
	\bibitem{EBTZinstability2019}
	W. Cong and R.B. Mann, \textit{Thermodynamic instabilities of generalized exotic BTZ black holes}, JHEP \textbf{11} (2019) 004.
	
		
	
\end{thebibliography}
\end{document}